\documentclass[prd,twocolumn,showpacs,superscriptaddress,floatfix]{revtex4}
\usepackage{epsfig}
\usepackage{amsmath}

\begin{document}

\title{Quantum Electrodynamics near a Dielectric Half-space}
\author{Claudia Eberlein}
\affiliation{Department of Physics \& Astronomy,
    University of Sussex, Falmer, Brighton BN1 9QH, England}
\author{Dieter Robaschik}
\affiliation{Lehrstuhl f\"ur Theoretische Physik,
Brandenburgische Technische Universit\"at Cottbus, Postfach 10 13 44, 
D--03013 Cottbus, Germany.}

\date{\today}
\begin{abstract}
We determine the photon propagator in the presence of a non-dispersive
dielectric half-space and use it to calculate the self-energy of an electron
near a dielectric surface.
\end{abstract}

\pacs{03.70}

\maketitle

\section{Introduction}
Quantum electrodynamics (QED) is liable to corrections if the
electromagnetic environment of the system under consideration is different
from free space. For example, the Lamb shift in an atom changes if the atom
is located not in free space but near a reflecting surface \cite{CP,hinds}. 
This and similar boundary-dependent effects are the subject of cavity QED
\cite{haroche}. Under most circumstances cavity QED effects are
non-relativistic in nature and hence the techniques employed in the theory
of cavity QED are chiefly non-relativistic. They rely mostly on a
comparatively simple mode expansion of the electromagnetic field and on
first-quantized theory for the remaining part of the system under
investigation (cf. e.g. \cite{BartonFawcett}). However, there are a few
examples of systems that are not inherently non-relativistic, the simplest
being a free electron. By virtue of being free it lacks an in-built energy
scale that could limit its virtual excitations and thus its interactions
with the electromagnetic field to non-relativistic energies. Other effects
that require a fully second-quantized theory are, e.g., radiative
corrections to the Casimir force between reflecting planes \cite{wiecz} and
the Scharnhorst effect of faster-than-$c$ light propagation inbetween and
perpendicular to parallel plates \cite{scharn}. To date all such
field-theoretical calculations have been done for cavity walls that are
perfectly reflecting. While it is obvious that no real material can ever
really be perfectly reflecting, the model of perfect reflectivity seems to
capture all the essential physics of the boundary without going wrong by
anything other than a minor numerical prefactor. Its great attraction is
that it is comparatively simple; for example, the photon propagator between
two parallel perfectly reflecting plates can be written as a sum of the
photon propagator in free space and a small boundary-dependent
correction, and loop calculations using it are manageable, though not trivial
because of the loss of translation invariance perpendicular to the plates
\cite{wiecz}.

However, we recently discovered that the assumption of perfect reflectivity
for the cavity walls is in fact not justified for systems, such as a free
electron, that admit low-frequency excitations, i.e. whose excitation
spectrum has, unlike an atom's, no natural IR cutoff \cite{PRL}. This is
because, crudely speaking, the electron's virtual excitations of arbitrarily
long wavelengths interact with evanescent electromagnetic field modes
originating inside a cavity wall, and the chief defect of the
perfect-reflector model is that it ignores all such evanescent modes. 
To show this we have modelled an imperfect reflector by a non-dispersive
dielectric and have calculated the self-energy of an electron in front of a
dielectric half-space. We have found that taking the limit of perfect
reflectivity in the result disagrees with the corresponding calculation that
assumes a perfectly reflecting wall from the outset. The two results differ
by a factor of 2 in one direction and even by sign in the other. While the
effect as such can already be seen in a non-relativistic calculation, we felt
that there was a need for a proper second-quantized calculation, mainly for
three reasons: {\em (i)} the non-relativistic calculation yields different
results for the two models but gives no clue as to the origin of this
discrepancy; {\em (ii)} there is nothing that {\em a priori} restricts the
electron's motion to non-relativistic energy scales and, in fact, the
interaction energy is being integrated up to infinity, which could
potentially lead to errors that pass by unnoticed in a purely
non-relativistic calculation; and {\em (iii)} there are other effects, as
e.g. the boundary-dependent $g-2$ correction to the electron's anomalous
magnetic moment, which might be affected and whose calculation requires
field-theoretical methods. Furthermore, in the face of a major discrepancy
it seems wise to check all possibilities.

\begin{figure}
\begin{center}
\centerline{\epsfig{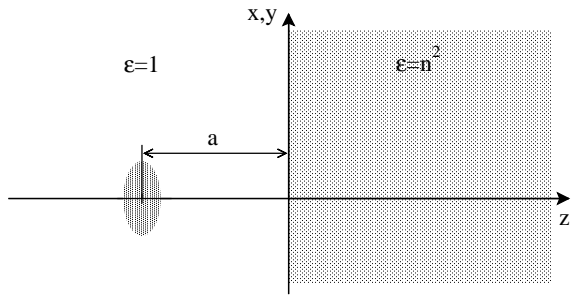}}
\end{center}
\caption{A sketch of the simplest system in cavity QED,
consisting of a quantum object, e.g. an atom or an electron, that is
located a distance $a$ away from a single reflecting wall. If the wall is
perfectly reflecting then the electromagnetic field is excluded from the
region $z>0$. A more realistic but still simple model is to describe the
wall by a constant, frequency-independent refractive index $n$. 
\label{fig:setup} }
\end{figure}

In this paper we establish the major building blocks of a full QED theory
near imperfect reflectors. We concentrate on a non-dispersive and
non-absorbing dielectric as a good model for an imperfectly reflecting
material. Thus the medium is characterized by a single parameter, its
refractive index $n$, which is real and the same for all frequencies. For
technical simplicity we restrict ourselves to a single reflecting surface,
i.e. we consider a dielectric half-space, which we take to occupy the region
$z>0$, while the region $z<0$ is vacuum (cf. Fig.~\ref{fig:setup}). For this
set-up the dielectric function is a single step function
$$
\epsilon({\bf r}) = 1 + \left(n^2-1\right) \theta(z)\;,
$$
which makes the solution of Maxwell's equations comparatively simple. For
piecewise constant dielectric functions like this, it is
advantageous gauge to work in the generalized Coloumb gauge
\begin{equation}
\pmb{\nabla}\cdot(\epsilon {\bf A}) = 0\;,
\label{eq:gaugecondition}
\end{equation}
which we shall do in this paper. For a general coordinate dependent
dielectric function $\epsilon({\bf r})$ the generalized Coulomb gauge may be
a very awkward choice, but for a piecewise constant $\epsilon({\bf r})$
this gauge is so convenient because
anywhere except right on the boundary (or boundaries) of the dielectric
($z=0$ in our case), it is equivalent to the Coulomb gauge $\pmb{\nabla} 
\cdot{\bf A} = 0$. Thus one can work almost as if in the Coulomb gauge and
just needs to make sure that the physical fields satisfy the appropriate
matching conditions at the boundary, i.e.~that
\begin{equation}
{\bf E}_\parallel\ \mbox{continuous}\ ,\ \ 
{\bf D}_\perp\ \mbox{continuous}\;.
\label{eq:cont}
\end{equation}
Since our model material is just a dielectric and has a magnetic
permeability $\mu=1$, the magnetic fields strengths ${\bf B}$ and ${\bf H}$
are also continuous everywhere. We note that quantum mechanics in the
generalized Coulomb gauge is related to that in the true Coulomb gauge by
canonical transformation \cite{barton1}; the two differ by surfaces charges
leading to an electrostatic image potential in the Hamiltonian.

If one wanted to work in a gauge that resembles the radiation
gauge, one ought to choose the gauge 
$$
\epsilon^2\frac{\partial\Phi}{\partial t} + \pmb{\nabla}\cdot(\epsilon
{\bf A})=0\;.
$$
In this gauge Maxwell's equations for the
scalar potential $\Phi$ and the vector potential ${\bf A}$ read
\begin{eqnarray}
&&\epsilon\frac{\partial^2\Phi}{\partial t^2} - \nabla^2\Phi 
= \frac{\rho}{\epsilon} + \frac{1}{\epsilon} (\pmb{\nabla}\Phi)\cdot
\pmb{\nabla}\epsilon\ ,\nonumber\\
&&\pmb{\nabla}(\pmb{\nabla}\cdot{\bf A})-\nabla^2{\bf A}+\epsilon 
\frac{\partial^2{\bf A}}{\partial t^2}-\epsilon\pmb{\nabla}\left[
\frac{1}{\epsilon^2}\pmb{\nabla}\cdot(\epsilon{\bf A})\right]={\bf j}\ .
\nonumber
\end{eqnarray}
It is important that they separate and that for piecewise constant
$\epsilon({\bf r})$ they differ from the standard wave equations for $\Phi$
and ${\bf A}$ only by surface terms. However, in the present paper we shall
not pursue any calculations in this gauge but work in the generalized Coulomb
gauge (\ref{eq:gaugecondition}).

In Section \ref{Sec:propagator} we calculate the full photon propagator in the
presence of a non-dispersive dielectric half-space starting from the normal
modes of the radiation field which we briefly discuss in Appendix
\ref{Sec:modes}. In Section \ref{Sec:self} we use this photon propagator
to determine the self-energy of a free electron located outside and a
distance $a$ away from the dielectric. In Section \ref{Sec:asympt} we do a
careful asymptotic analysis of the expression for the self-energy for
non-relativistic mean energies. In Section \ref{sec:ninf} we
compare the results for the electron's radiative self-energy in front of an
imperfectly or of a perfectly reflecting surface and discuss the reasons for
the disagreement between the calculation for a non-dispersive dielectric and
for a ``perfect reflector''. Section \ref{Sec:final} summarizes our final
results. 

Throughout this paper we set $c=1=\hbar$ and use Heaviside-Lorentz units for
electromagnetic quantities, $\epsilon_0=1=\mu_0$. Thus the fine structure
constant is $e^2/4\pi\approx 1/137$.

\section{Calculation of the photon propagator\label{Sec:propagator}}
\subsection{Wightman function}\label{Sec:Wightman}
\begin{figure}
\begin{center}
\centerline{\epsfig{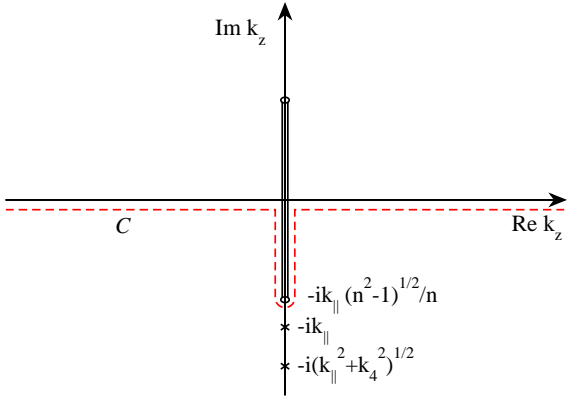}}
\end{center}
\caption{The integration  path $\mathcal{C}$ in the complex $k_z$ plane.
\label{fig:contour} }
\end{figure}

The Green's functions in free quantum electrodynamics are
vacuum expectation values of products of field operators. Let us first
consider the Wightman functions \cite{bog} 
\begin{eqnarray}
\label{wd0}
D^{\mu\,\nu\,-}(x,x')= -i<0|A^\mu(x) A^\nu(x')|0>\;. 
\end{eqnarray}
Inserting the normal modes (\ref{left}) and (\ref{right}) into (\ref{wd0})
and taking the vacuum expectations values of the bilinear products of
photon annihilation and creation operators, we obtain
\begin{eqnarray}
\label{wd1}
D^{\mu\,\nu\,-}(x,x')= \sum_\sigma e^{\mu}_\sigma(\partial_x)\, 
            e^{\nu *}_\sigma(\partial_{x'})\; g^{\sigma \sigma}
\hspace*{15mm}\nonumber\\
            \times\big[{^LD}_\sigma^-(x,x')+ {^RD}_\sigma^-(x,x')\big].
\end{eqnarray}
with
\begin{widetext}
\begin{eqnarray}
{^RD}_\sigma^-(x,x') 
&=&i\int \frac{d^2k_\parallel}{(2\pi)^3} \int_{-\infty}^0 dk^d_z\;
\frac{1}{2\omega} \    e^{-i(\tilde x-\tilde x')\tilde k}\Big\{  
\theta(-z)\theta(-z') \big[
                 e^{(ik_zz-ik_z^*z')}T^R_\sigma {T^R_\sigma}^*   
              \big] \nonumber \\
&&+ \theta(z)\theta(z')\frac{1}{n^2}\big[
   e^{ik_z^d(z-z')} 
                     +e^{ik^d_z(z+z')} {R^ R_\sigma (k)}^*
                     +e^{-ik_z^d(z+z')} R^R_\sigma(k)
                     +e^{i(k_z^d (z -z')}R^R_\sigma {R^R_\sigma}^* 
                       \big]  
                     \nonumber \\     
&&+\theta(-z)\theta(z')\frac{1}{n}\big[
      e^{(ik_zz-ik_z^d z')} T^R_\sigma
       + e^{ik_zz+ik_z^d z')}T^R_\sigma {R^R_\sigma}^* 
     \big] \nonumber \\
&&+\theta(z)\theta(-z')\frac{1}{n}\big[ 
     e^{(ik_z^dz-ik_z^* z')} {T^R_\sigma}^*
  +  e^{-ik_z^dz-ik_z^* z')} R^R_\sigma {T^R_\sigma}^* 
                    \big]\Big \}\;, \label{wd2r}
\end{eqnarray}
and
\begin{eqnarray}
{^LD}_\sigma^-(x,x')
&=&i\int \frac{d^2k_\parallel}{(2\pi)^3} \int_0^\infty d k_z\;
\frac{1}{2\omega} \    e^{-i(\tilde x-\tilde x')\tilde k}\Big\{ 
\theta(z)\theta(z')\frac{1}{n^2}   e^{ik_z^d(z-z')}
           T^L_\sigma{T^L_\sigma}^*  \nonumber \\
&&+\theta(-z)\theta(-z') \big[
             e^{ik_z(z-z')}  
              +e^{ik_z(z+z')} {R^L_\sigma(k)}^*
             +e^{-ik_z(z+z')} R^L_\sigma(k)
             + e^{-ik_z(z-z')} R^L_\sigma(k){R^L_\sigma(k)}^* 
                 \big] \nonumber \\
& &+\theta(-z)\theta(z')\frac{1}{n}\big[
   e^{(ik_zz-ik_z^d z')} {T^L_\sigma}^*
                + e^{(-ik_zz-ik_z^d z')} {T^L_\sigma}^* R^L_\sigma \big] 
                \nonumber \\
& &+\theta(z)\theta(-z')\frac{1}{n}\big[ 
           e^{(ik_z^dz-ik_z z')} T^L_\sigma +
              e^{(ik_z^d z+ik_z z')} T^L_\sigma {R^L_\sigma}^* 
         \big]\Big\}\;. \label{wd2l} 
\end{eqnarray}
The notations for the wave vectors is as defined in Eq.~(\ref{k}), and
in addition we have introduced the new variables $\tilde k = (k^0,{\bf
k}_\parallel)$ and $\tilde x = (x^0,{\bf x}_\parallel)$ in 2+1 dimensional
Minkowski space.  The sum of ${^LD}_\sigma^-(x,x')$ and 
${^RD}_\sigma^-(x,x')$ can be simplified by taking into account that the
Fresnel coefficients are real functions of the wave vectors and by using
various relations (\ref{eq:relreal}) and (\ref{eq:relimag}) between them and
their products. We obtain two equivalent expressions:
\begin{eqnarray}
{^LD}_\sigma^-(x,x')+{^RD}_\sigma^-(x,x')&=& 
i\int \frac{d^2k_\parallel}{(2\pi)^3} \frac{1}{2\omega}\ 
     e^{-i(\tilde x-\tilde x')\tilde k}\Bigg \{ 
\theta(z)\theta(z')\frac{1}{n^2}
   \int_{-\infty}^\infty dk_z^d \left[ e^{ik_z^d(z-z')} 
                     +e^{-ik_z^d(z+z')} R^R_\sigma \right] \nonumber \\
& &+\theta(-z)\theta(-z') \left[
   \int_{-\infty}^{\infty} dk_z e^{ik_z(z-z')} 
   +\int_{-\infty}^{\infty} dk_z e^{-ik_z(z+z')} R^L_\sigma
   +\int_{-\Gamma}^0 dk_z^d e^{ik_zz-ik_z^*z'} {T^R_\sigma}^*T^R_\sigma  
       \right] \nonumber \\
& &+\theta(-z)\theta(z')\frac{1}{n}
   \int_{-\infty}^\infty dk_z^d  e^{ik_zz-ik_z^d z'} T^R_\sigma
   +\theta(z)\theta(-z')\frac{1}{n}
   \int_{-\infty}^\infty dk_z  e^{-ik_zz'+ik_z^d z} {T^R_\sigma}^*
   \Bigg\} \label{wd4}
\end{eqnarray} 
and
\begin{eqnarray}
{^LD}_\sigma^-(x,x')+{^RD}_\sigma^-(x,x')&=& 
i\int \frac{d^2k_\parallel}{(2\pi)^3} \frac{1}{2\omega}\ 
     e^{-i(\tilde x-\tilde x')\tilde k}\Bigg \{     
\theta(z)\theta(z')\frac{1}{n^2}
   \int_{\infty}^\infty dk_z^d \left[ e^{ik_z^d(z-z')} 
                     +e^{-ik_z^d(z+z')} R^R_\sigma \right] \nonumber \\
&&+     \theta(-z)\theta(-z') \left[
   \int_{-\infty}^{\infty} dk_z e^{ik_z(z-z')} 
   + \int_\mathcal{C} dk_z e^{ik_z(z+z')} R^L_\sigma \right] \nonumber \\
& &+\theta(-z)\theta(z')\frac{1}{n}
   \int_{-\infty}^\infty dk_z^d  e^{ik_zz-k_z^d z'} T^R_\sigma
   +\theta(z)\theta(-z')\frac{1}{n}
   \int_{-\infty}^\infty dk_z  e^{-ik_zz'+ik_z^d z} {T^R_\sigma}^*
    \Bigg \}, \label{wd3}
\end{eqnarray} 
\end{widetext}
with $\Gamma= ((n^2-1)k_\parallel^2)^{1/2}$. The difference between these
two expressions is how the contributions from evanescent waves are
included. In the first expression (\ref{wd4}) they appear in the second line
as a separate integral over $k_z^d$ from $-\Gamma$ to 0. In the second
expression (\ref{wd3}) they are included in the integration along the path
$\mathcal{C}$ in the complex $k_z$ plane shown in Fig.~\ref{fig:contour}: it
runs along the real axis from $-\infty $ to $0$, then down the negative
imaginary axis from $0$ to $-i\Gamma/n$ to the left of the square root cut,
back up to the origin to the right of the cut, and then along the real axis
from $0$ to $+\infty$. The cut is due to $ k_z^d = (n^2 k_z^2 +
(n^2-1)k_\parallel^2)^{1/2}/n $ and extends from $k_z = + i\Gamma/n$ to $k_z
= - i\Gamma/n$. The part of $\mathcal{C}$ that runs left and right of the
cut is identical to the integral over $k_z^d$ in the second line of
(\ref{wd4}), i.e.~it gives the contribution of the of the evanescent waves.
This works because:
\begin{eqnarray}
\left.R^L\right|_{k_z=-i\kappa,\,k_z^d =-K}- 
 \left.R^L\right|_{k_z=-i\kappa,\,  k_z^d =K}\hspace*{15mm}\nonumber\\ 
= \left.\frac{k_z}{k_z^d}n^2 T^R {T^R}^*\right|_{k_z=-i\kappa,\,
    k_z^d= -K}\;.
\end{eqnarray}
Note that the $\omega =0$ singularity at $k_z= \pm i k_\parallel $
does not come into play if one chooses the cut of $\omega = (k_z^2 +
k_\parallel^2)^{1/2}$ along $-i \infty\ldots -ik_\parallel$ and
$ik_\parallel\ldots i \infty$.

\subsection{Feynman propagator}

For the calculation of radiative corrections we need the Feynman propagator,
which can be reconstructed from the Wightman functions (\ref{wd1}), (\ref{wd4})
and (\ref{wd3}) according to
\begin{eqnarray}
\label{fd0}
D^{\mu\,\nu\,c}(x,x')&=& -i\langle 0|TA^\mu(x) A^\nu(x')|0\rangle \nonumber \\ 
                     &=&\theta(x^0 -{x'}^0)\; D^{\mu\,\nu\,-}(x, x')
\nonumber\\&&
        +\theta({x'}^0-x^0)\;D^{\nu\,\mu\,-}(x', x) \nonumber
\end{eqnarray}
Thus $D^{\mu\,\nu\,c}(x,x')$ can be written in the same way as the Wightman
function in (\ref{wd1}),
\begin{displaymath}
D^{\mu\,\nu\,c}(x,x') = \sum_\sigma e^{\mu}_\sigma(\partial_x)\,
            e^{\nu *}_\sigma(\partial_{x'})\; g^{\sigma \sigma}\;
	    D^{c}_\sigma(x,x')\;.
\end{displaymath}
The sum over the polarizations is gauge dependent. In Coulomb gauge the sum
runs over $\sigma = TE,\ TM$ only. In covariant, i.e.  Feynman gauge the two
unphysical polarizations $G,\ C$ have to be included.  Proceeding from the
simplified expressions (\ref{wd4}) and (\ref{wd3}) for the Wightman
functions, we obtain for the polarization component $\sigma$ of the Feynman 
propagator
\begin{widetext}
\begin{eqnarray}
\label{fd2}
D_\sigma^c(x,x')&\!\!\! = &\!\!\!
-\!\int\! \frac{d^3 \tilde k}{(2\pi)^4}\;
     e^{-i(\tilde x-\tilde x')\tilde k} \left\{\; \theta(z)\theta(z')
   \int_{-\infty}^{\infty} dk_z^d \left[ e^{ik_z^d(z-z')} +
   e^{-ik_z^d(z+z')}R^R_\sigma \right] 
\frac{1}{n^2 k_0^2 - k_p^2 -k_z^{d\,2} +i\epsilon} \right. \\
& &+\theta(-z)\theta(-z') \left[
   \int_{-\infty}^{\infty} dk_z\;  e^{ik_z(z-z')} +
\int_{\cal C} dk_z  e^{ik_z(z+z')}R^L_\sigma \right] \frac{1}{k^2 +i\epsilon} 
\nonumber \\
& &+\theta(-z)\theta(z')\frac{1}{n} 
   \int_{-\infty}^{\infty} dk_z^d\;  e^{ik_z z -ik_z^d z'}\;T^R_\sigma \;  
    \frac{1}{k_0^2 -(k_p^2+k_z^{d\,2})/n^2 +i\epsilon}
\nonumber \\
& &\left. +\theta(z)\theta(-z')\frac{1}{n} 
   \int_{-\infty}^{\infty} dk_z^d\; e^{-ik_zz' +ik_z^d z}\;T^{R*}_\sigma \;  
    \frac{1}{k_0^2 -(k_p^2+k_z^{d\,2})/n^2 +i\epsilon}
    \right\} \;.\nonumber
\end{eqnarray}
In an alternative formulation one can replace the second line of (\ref{fd2})
by
\begin{equation}
\label{fd3}
\theta(-z)\theta(-z') \left[
   \int_{-\infty}^{\infty} dk_z \; e^{ik_z(z-z')} +
\int_{-\infty}^\infty dk_z\;  e^{ik_z(z+z')}R^L_\sigma
        +\int_{-\Gamma}^0 dk_z^d\; e^{ik_zz-ik^*_zz'}
         T^{R*}_\sigma\;T^R_\sigma
         \right] \frac{1}{k^2 +i\epsilon}
\end{equation}
\end{widetext}
Note that the property $ e^\mu_\lambda(\partial_x)
e^{\nu *}_\lambda(\partial_{x'})= 
e^{\mu *}_\lambda(\partial_x) e^\nu_\lambda(\partial_{x'}) $ follows
directly from the definition (\ref{eq:polvecs}) of the polarization vectors.
Note also that $k_0$ in Eq.~(\ref{fd2}) is a free integration variable and
is not fixed to $\omega$ as it was in the case of the Wightman functions
(\ref{wd4}) and (\ref{wd3}).

The propagator differs from its equivalent in free space by not being
translation invariant along $z$. However, it is still symmetric in its
arguments, $D^c_\sigma(x,x')=D^c_\sigma(x',x)$. This is easily shown by a
change of variables $ k \rightarrow -k$. For real $k_z$ this is
straightforward; for imaginary $k_z = -i\kappa$ (with $\kappa>0$)
one needs the relation
\begin{eqnarray*}
T^R (k_z= -i\kappa, -k_z^d) = {T^{R*}(k_z= -i\kappa, k_z^d)}\;,
\end{eqnarray*}
for which one has to take care to stay on the same sheet in the complex plane. 

The wave equations that the Feynman propagator satisfies are
\begin{eqnarray*}
 (\partial_{x_0}^2 - \Delta_x)D^c_\sigma(x,x') &=& \delta^{(4)}(x-x')\;,
 \qquad\mbox{for }  z<0\;, \\
 (n^2 \partial_{x_0}^2 -  \Delta_x)D^c_\sigma(x,x') &=& \delta^{(4)}(x-x')\;,
 \qquad\mbox{for }  z>0\;.
\end{eqnarray*}
Checking the wave equation for example for $z<0$ leads to
\begin{widetext}
\begin{eqnarray*}
  (\partial_{x_0}^2 - \Delta_x) D_\sigma^c(x,x') &\!\!\!=&\!\!\! 
\theta(-z)\theta(-z') \delta^{(4)}(x-x')\\ && \hspace*{-20mm}
   +\int \frac{d^3 \tilde k}{(2\pi)^4} e^{-i(\tilde x-\tilde x')\tilde k}
 \left\{ \theta(-z)\theta(-z') 
     \int_{-\infty}^{\infty} dk_z\;  e^{ik_z(z-z')}R^L_\sigma  
     + \Theta(-z)\Theta(z')\frac{1}{n} 
   \int_{-\infty}^{\infty} dk_z^d \; e^{ik_z z -ik_z^d z'}T^R_\sigma \right\}. 
\end{eqnarray*}
\end{widetext}
The second line should vanish, which it does --- because the integration path
can be closed in the lower $ k_z$ plane as the Fresnel coefficients do not
have poles there.

A direct check of the Feynman propagator is the reconstruction of the
Wightman functions (\ref{wd4}) and (\ref{wd3}) by performing the $k_0$
integration. For $x_0 >0$ the integration path can be closed in the lower
half plane
\begin{eqnarray}
\label{exp}
 \int_{-\infty}^\infty \frac{\exp(-ik_0x^0)}{k_0^2 + {\bf k}^2 +i\epsilon }
  = -\frac{i}{2|{\bf k}|}\;
         \exp(-i|{\bf k}|x^0)\; \theta(x_0) \;,
\end{eqnarray}
and we thus reproduce (\ref{wd4}) and (\ref{wd3}) for $x_0 -{x'}_0 >0 $.\\

Next we consider two special cases. The simplest situation is the limit
$n \rightarrow 1 $, when there is no medium. Correspondingly
all reflection coefficients $R$ tend to zero, the transmission 
coefficients $T$ approach one, and $\Gamma\rightarrow 0$, so that complex
values do not appear in the integration over $k_z$.
For all polarizations $\sigma$ the propagator components equal the massless 
scalar propagator
\begin{eqnarray}
\label{n1}  
 D_\sigma^c(x,x') = 
            -\int \frac{d^4 k}{(2\pi)^4}
            e^{-i( x- x') k} \frac{1}{k^2 +i\epsilon} \;.
\end{eqnarray}
Thus, according to (\ref{metrel}) our Feynman propagator (\ref{fd2})
coincides in the limit $n \rightarrow 1 $ with the standard free space
propagators: either with the covariant propagator in Feynman gauge or with
the propagator in Coulomb gauge, depending on which modes have been included
in the sum over polarizations.
 
In the limit $ n \rightarrow \infty $ only the left incident modes survive,
so that, according to (\ref{inft}), the Wightman functions (\ref{wd1})and
its components (\ref{wd4}) simplify greatly.
\begin{eqnarray}
\label{intfw}
{D}_\sigma^-(x,x')
=i\int \frac{d^2k_\parallel}{(2\pi)^3} \int_{-\infty}^\infty \frac{d k_z}{2\omega}
     e^{-i(\tilde x-\tilde x')\tilde k} \theta(-z)\theta(-z') \nonumber \\ 
\times     
\left[e^{ik_z(z-z')} +e^{ik_z(z+z')} {R^L_\sigma(k)} \right] \hspace*{10mm}
\end{eqnarray}
with 
\[ R^L_{TE} = R^L_{C} =R^L_{G} = -1, R^L_{TM} = 1. \] 
A corresponding representation follows for the Feynman propagator.

Finally we need to establish the electrostatic Green function, which corresponds
to the Coulomb interaction. 
One way to start is from the retarded Green function for the Coulomb mode
$C$. The retarded propagator differs from the Feynman propagator (\ref{fd2}) by the 
$i\epsilon$ prescription in the denominator, such that 
$i\epsilon \rightarrow i\epsilon k_0 $. Since we are looking for a static
Green's function we need to calculate
\begin{widetext}
\begin{eqnarray}\label{greenc}
  G_C(\vec x, \vec x') &=& \int dx^0\; D_C^{ret}(x,x') \\
                       &=&  \int \frac{d^2 k_\parallel}{(2\pi)^3}
e^{i({\bf x}_\parallel -{\bf x}'_\parallel)\cdot{\bf k}_\parallel}
\frac{1}{{\bf k}^2}
\left\{ \theta(-z)\;\theta(-z') \int_{-\infty}^{\infty} dk_z \left( e^{ik_z(z-z')} +
e^{ik_z(z+z')}\frac{1-n^2}{1+n^2}\right)\right. \nonumber \\
& &+\;\theta(z)\;\theta(z')
   \int_{-\infty}^{\infty} dk_z \left( e^{ik_z(z-z')} +
 e^{-ik_z(z+z')}\frac{n^2-1}{n^2 +1}\right) 
           \nonumber \\ 
& &\left.+\;\theta(-z)\;\theta(z') 
   \int_{-\infty}^{\infty} dk_z \; e^{ik_z (z- z')} \frac{2n}{1+n^2}
   +\;\theta(z)\;\theta(-z') 
   \int_{-\infty}^{\infty} dk_z \; e^{ik_z(z - z')}
 {\frac{2n}{1+n^2}}
     \right\}\;. \nonumber
\end{eqnarray}
In our conventions $A^0(x)=\Phi^0(x)$ for $z<0$ and $A^0(x)=n\Phi^0(x)$ for $z>0$,
so that we obtain for the electrostatic Green's function 
\begin{eqnarray}
 G_{\Phi}(x,x') &=&
-i\langle 0|\Phi^0(x) \Phi^0(x')|0\rangle \nonumber \\
 &=&  \int \frac{d^2 k_\parallel}{(2\pi)^3} \int_{-\infty}^{\infty} dk_z\; 
e^{i({\bf x}_\parallel -{\bf x}'_\parallel)\cdot{\bf k}_\parallel}
\frac{1}{{\bf k}^2}
\left\{ \theta(-z)\;\theta(-z') \left( e^{ik_z(z-z')} +
e^{ik_z(z+z')}\frac{1-n^2}{1+n^2}\right)\right. \label{greencphi}\\
& &+\;\theta(z)\;\theta(z')\;\frac{1}{n^2}\left( e^{ik_z(z-z')} +
 e^{-ik_z(z+z')}\frac{n^2-1}{n^2 +1}\right) 
           \nonumber \\ 
& &\left.+\;\theta(-z)\;\theta(z')  \; e^{ik_z (z- z')} \frac{2}{1+n^2}
   +\;\theta(z)\;\theta(-z')  \; e^{ik_z(z - z')}
 {\frac{2}{1+n^2}} 
     \right\}\;. \nonumber
\end{eqnarray}
\end{widetext}
This result agrees with the classical electrostatic Green's function, which
can be derived easily, so that we have an additional check.

\section{The self-energy of the electron\label{Sec:self}} 
The energy shift of the electron can be determined by considering the
electron propagator and its radiative corrections due to the coupling to the
photon field \cite{Weinberg}. We start be considering the free electron propagator
\begin{equation}
S_{\alpha\beta}^{(0)}(x,y) = \langle 0 |\, {\sf T}\, \psi_{\alpha}(x)\, 
\overline{\psi}_\beta(y) |0\rangle \;.
\label{eq:S0}
\end{equation}
Using canonical quantization one can write the spinor
\begin{eqnarray}
\psi_{\alpha}({\bf r}, t) = \int\! \frac{d^3{\bf p}}{(2\pi)^{3/2}}
\sqrt{\frac{m}{E}}
\sum_{i=1,2} \left[ b_{i,{\bf p}} u_\alpha^{(i)}({\bf p}) 
e^{-iEt+i{\bf p}\cdot{\bf r}}\right.\nonumber\\\left.
+ d_{i,{\bf p}}^\dagger v_\alpha^{(i)}({\bf p}) 
e^{iEt-i{\bf p}\cdot{\bf r}}\right]\;,
\hspace*{5mm}\label{eq:spinorexpansion}
\end{eqnarray}
where $b_{i,{\bf p}}$ annihilates an electron of helicity $i$ and momentum
${\bf p}$ and $d_{i,{\bf p}}^\dagger$ creates a positron of helicity $i$ and momentum
${\bf p}$. The particle eigenspinors $u_\alpha^{(i)}({\bf p})$ are
solutions of the Dirac equation, 
\begin{eqnarray}
\label{h3}
       u^{(i)}({\bf p})= \sqrt{\frac{p^0 +m}{2m}}
       \left(\begin{array}{c}
         \phi^{(i)}\\ \\
         \displaystyle\frac{\boldsymbol{\sigma}\cdot{\bf p}}{p_0+m}\;\phi^{(i)} 
       \end{array} \right),
\label{eq:eigenspinors}
\end{eqnarray}
with $\phi^{(i)}$ two orthogonal and normalized two-spinors. Thus the
normalization of $u^{(i)}({\bf p})$ is 
\[
\overline{u}^{(i)}({\bf p})\, u^{(j)}({\bf p})= \delta_{ij}\,.
\]
The antiparticle eigenspinors $v^{(i)}({\bf p})$ are similar to (\ref{h3}),
except with upper and lower components interchanged and
normalized to -1. Inserting the canonical expansion
(\ref{eq:spinorexpansion}) and its conjugate into the expression for the
propagator (\ref{eq:S0}) and Fourier transforming to go from the time variable
$(x_0-y_0)$ to the energy $E$ one finds
\begin{widetext}
\begin{eqnarray}
S_{\alpha\beta}^{(0)}({\bf x},{\bf y}, E)= 
- \int\! \frac{d^3{\bf p}}{(2\pi)^3} \frac{m}{E({\bf p})} \sum_{i=1,2} 
\frac{u^{(i)}({\bf p})\overline{u}^{(i)}({\bf p})}{E({\bf p})-E-i\epsilon} 
e^{i{\bf p}\cdot({\bf x}-{\bf y})}
+\int\! \frac{d^3{\bf p}}{(2\pi)^3} \frac{m}{E({\bf p})} \sum_{i=1,2} 
\frac{\overline{v}^{(i)}({\bf p})v^{(i)}({\bf p})}{E({\bf p})+E-i\epsilon} 
e^{-i{\bf p}\cdot({\bf x}-{\bf y})}\;.
\hspace*{5mm}
\label{eq:S0energy}
\end{eqnarray}
This shows that the propagator has positive energy poles at the particle
energies $E({\bf p})$ and negative energy poles at the antiparticle energies
$-E({\bf p})$. Radiative corrections represent a perturbation and cause
shifts in the particle and antiparticles eigenfunctions and in their
energies. For small perturbations these shifts are small and expressions can
be
linearized in them. Linearizing the change of the propagator $\delta
S=S-S^{(0)}$, one obtains a
term that is linear in the energy shift and has a double pole at the
particle or antiparticle energy,
\begin{eqnarray}
\delta S_{\alpha\beta}^{(0)}({\bf x},{\bf y}, E) &\!\simeq &\!\int\! \frac{d^3{\bf p}}{(2\pi)^3} \frac{m}{E({\bf p})} \sum_{i=1,2}
\left\{ \frac{u^{(i)}({\bf p})\overline{u}^{(i)}({\bf p})}{\left[
E({\bf p})-E-i\epsilon\right]^2} 
e^{i{\bf p}\cdot({\bf x}-{\bf y})}
- \frac{\overline{v}^{(i)}({\bf p})v^{(i)}({\bf p})}{\left[
E({\bf p})+E-i\epsilon\right]^2} 
e^{-i{\bf p}\cdot({\bf x}-{\bf y})}\right\}\,\delta E({\bf p})
\nonumber\\
&&+ \ \mbox{(other terms without double poles)}\;.
\label{eq:deltaS}
\end{eqnarray}
\end{widetext}
Further terms that are linear in the shifts of the particle and antiparticle
eigenfunctions do not give rise to terms with double poles since the
eigenfunctions appear only in the numerator of Eq.~(\ref{eq:S0energy}).

The energy shift $\delta E({\bf p})$ can now be determined by comparing the
expression (\ref{eq:deltaS}) to the change of the electron propagator as
determined from standard Feynman perturbation theory. At one-loop level,
i.e. to order $e^2$, the radiative correction to the propagator is
\begin{eqnarray}
\delta S_{\alpha\beta}^{(0)}({\bf x},{\bf y}, E) = \int d^3{\bf z}
\int d^3{\bf z}'\; S^{(0)}_{\alpha\gamma}({\bf x},{\bf z},E) 
\Sigma_{\gamma\kappa}({\bf z},{\bf z}',E)
\nonumber\\
\times\ S^{(0)}_{\kappa\beta}({\bf z}',{\bf y},E)\;,\hspace*{5mm}
\label{eq:oneloop}
\end{eqnarray}
where $\Sigma_{\alpha\beta}({\bf x},{\bf x}',E)$ is obtained from the 
standard electron self-energy
\pagebreak[3]
\begin{eqnarray}
\label{self0}
\Sigma_{\alpha\beta}(x,x') = - i e^2\gamma^\mu_{\alpha\kappa} 
S^{(0)}_{\kappa\lambda}(x-x')
\gamma^\nu_{\lambda\beta} D^c_{\mu \nu}(x,x')\;
\end{eqnarray}
by Fourier transformation from the time into the energy domain.
The electron propagator $S^{(0)}(x-x')$ is the same as in free space and
thus translation invariant in all 4 directions, but the photon propagator
$D^c_{\mu \nu}(x,x')$ is affected by the presence of the dielectric medium
and therefore not translation invariant in the $x_3$ direction. Substituting
the representation (\ref{eq:S0energy}) in terms of eigenfunctions into 
Eq.~(\ref{eq:oneloop}) one obtains
\begin{widetext}
\begin{eqnarray}
\delta S_{\alpha\beta}^{(0)}({\bf x},{\bf y}, E) = \int d^3{\bf z}
\int d^3{\bf z}' \int\! \frac{d^3{\bf p}}{(2\pi)^3} \int\! 
\frac{d^3{\bf p}'}{(2\pi)^3}  \frac{m}{E({\bf p})} \frac{m}{E({\bf p}')} 
\sum_{i=1,2} \sum_{j=1,2} \left[ e^{i{\bf p}\cdot({\bf x}-{\bf z}) +
i{\bf p}'\cdot({\bf z}'-{\bf y})} 
\frac{u^{(i)}_\alpha({\bf p})\overline{u}^{(i)}_\gamma({\bf p})}{
E({\bf p})-E-i\epsilon}\right.\nonumber\\ \left. \times \ 
\Sigma_{\gamma\kappa}({\bf z},{\bf z}',E)
\frac{u^{(j)}_\kappa({\bf p}')\overline{u}^{(j)}_\beta({\bf p}')}{
E({\bf p}')-E-i\epsilon}
+ \ldots \right]
\label{eq:deltaSloop}
\end{eqnarray}
\end{widetext}
Further terms all contain antiparticle operators and at least one negative
energy pole. Since we are interested in the energy shift of a particle
rather than an antiparticle, we need to focus only on terms with two
particle poles. For ${\bf p}={\bf p}'$ Eq. (\ref{eq:deltaSloop}) has the
same double pole as Eq. (\ref{eq:deltaS}), and thus a simple comparison of
the coefficients of those double-pole terms should yield an expression for
the energy shift $\delta E({\bf p})$ in (\ref{eq:deltaS}). However, a
mathematically clean comparison is possible only if one introduces a
quantization volume $L^3$ with periodic boundary conditions so as to
discretize the momentum ${\bf p}$. Then all integrals over momenta turn into
sums according to the prescription
\[
\int\! \frac{d^3{\bf p}}{(2\pi)^3} \longrightarrow \frac{1}{L^3} \sum_{\bf p}\;.
\]
Only the term with ${\bf p}={\bf p}'$ in the double sum over momenta in
Eq. (\ref{eq:deltaSloop}) gives rise to a double pole in the energy, and
comparison with the double-pole term in Eq. (\ref{eq:deltaS}) therefore yields 
\begin{eqnarray}
\delta E({\bf p}) = \frac{1}{L^3} \int d^3{\bf z}\int d^3{\bf z}' 
\frac{m}{E({\bf p})}\, e^{-i{\bf p}\cdot({\bf z}-{\bf z}')}\,
\overline{u}^{(i)}_\gamma({\bf p}) \nonumber\\ \times \ 
\Sigma_{\gamma\kappa}({\bf z},{\bf z}',E)\,
u^{(i)}_\kappa({\bf p})\;.
\label{eq:dEzz}
\end{eqnarray}
It is advantageous to work with the Fourier representation of $\Sigma$.
Because of the lack of translation invariance in the $x_3$ direction, the
Fourier transform of the self energy with respect
to ${\bf x}-{\bf x}'$ has a residual dependence on $x_3+x_3'$,
\begin{eqnarray}
\label{h6}
 \Sigma_{\alpha\beta}(x, x')= \int \frac{d^4q}{(2\pi)^4}\; e^{-iq(x -x')}\; 
           \Sigma_{\alpha\beta} (q,x_3+x'_3)\;.
\end{eqnarray}
In a finite quantization volume the integral over $d^3{\bf q}$ again turns into a
sum, and we can re-write Eq.~(\ref{eq:dEzz}) as
\begin{eqnarray}
\delta E({\bf p}) = \frac{1}{L^6} \int d^3{\bf z}\int d^3{\bf z}' \sum_{\bf q}
\frac{m}{E({\bf p})}\, e^{i({\bf q}-{\bf p})\cdot({\bf z}-{\bf z}')}\,
\overline{u}^{(i)}_\gamma({\bf p}) \nonumber\\ \times \ 
\Sigma_{\gamma\kappa}(q,z_3+z_3')\,
u^{(i)}_\kappa({\bf p})\;.\hspace*{5mm}
\label{eq:dEdiscrete}
\end{eqnarray}
Since we want to work out the energy shift of a particle as a function of
its distance from the dielectric, we need to form localized wave packets in
the $x_3$ direction --- so that the concept of a certain distance between
the electron and the surface of the dielectric at all makes sense. If the
centre of the packet is at $x_3=-a$ (see Fig.~\ref{fig:setup}) then we can
approximate $z_3+z_3'\simeq -2a$ and carry out the ${\bf z}$ and ${\bf z}'$ 
integrations in (\ref{eq:dEdiscrete}). The result simplifies to
\begin{equation}
\delta E({\bf p}) = \frac{m}{E({\bf p})}\,
\overline{u}^{(i)}_\gamma({\bf p})\,
\Sigma_{\gamma\kappa}(p,-2a)\, u^{(i)}_\kappa({\bf p})\;.
\label{eq:dE}
\end{equation}
Note that, while $\Sigma (q,x_3+x'_3)$ in Eq.~(\ref{h6})
is in general off shell, it is on
the mass shell in Eqs.~(\ref{eq:dEdiscrete}) and (\ref{eq:dE}) because
$q$ and $p$ are on shell.

Further we need
to remark that in Coulomb gauge the energy shift is not wholly due to the
radiative self energy (\ref{self0}): we have to add to (\ref{eq:dE}) the
electrostatic energy
\begin{eqnarray}
\label{h1c}
    \delta E _{\mbox{\scriptsize Coulomb}} = 
          \frac{e^2}{2}\int d^3x \int d^3x'\;\langle
          :\overline{\psi}(x) \gamma^0 \psi(x): \nonumber\\
\times\ \overline{G}_\Phi(x,x')\;
          :\overline{\psi}(x') \gamma^0 \psi(x'):\rangle\,\,,  
\end{eqnarray}
where $\overline{G}_\Phi$ is the part of the electrostatic Green's function
(\ref{greencphi}) that depends on the presence of the dielectric.
The electrostatic shift is easy to evaluate, which we shall do in Section 
\ref{Sec:elstatic}.

We are interested in the self-energy corrections for an electron located
well outside the dielectric. Because of the electron's localization its
direct interaction with the dielectric medium is completely negligible,
i.e. there is no wave-function overlap between the electron and the
microscopic constituents of the dielectric. That is why, for $x_3<0$ and 
$x'_3<0$ we can work with the standard free electron propagator,
\begin{eqnarray}
\label{elepr} 
S^{(0)}(x-x')&=& (i \gamma^\mu \partial_{x^\mu} + m) D^c_m(x-x')\;, \\
D^c_m(x-x')&=&  -\int \frac{d^4q}{(2\pi)^4}\, e^{-iq(x-x')}\,
              \frac{1}{q^2 -m^2 + i \epsilon}\;. \nonumber        
\end{eqnarray}
The impact of the dielectric medium onto the self-energy of the electron is
consequently just due to the electromagnetic interaction, i.e. due to the
fact that the photon propagator (\ref{fd2}) depends on the presence and
electromagnetic properties of the medium. Since we are interested only in
the energy shift due to the presence of the dielectric, we split the photon
propagator into the free photon propagator and a medium-dependent part,
\[D^c(x,x')=D^c_{(0)}(x,x') + \overline{D}(x,x')\;,
\]
and take only the medium-dependent part $\overline{D}(x,x')$ for calculating
the self-energy (\ref{self0}) and the energy shift (\ref{eq:dE}). This also
means that we do not have to deal with regularization and renormalization;
these have been done in the free part of the photon field, and we work with
already renormalized quantities. All medium-dependent corrections will then
automatically be finite.

From (\ref{fd2}) we see that for $z<0$ and $z'<0$, i.e. outside the
dielectric, the medium-dependent part of the photon propagator is
\begin{eqnarray}
&&\hspace*{-5mm}
\overline{D}^{\mu\,\nu\,c}(x,x')|_{z,z'<0} = -\sum_\sigma e^{\mu}_\sigma(\partial_x)\,
            e^{\nu *}_\sigma(\partial_{x'})\, g^{\sigma \sigma}
\overline{D}^{c}_\sigma(x,x')\nonumber\\
&&\hspace*{-5mm}\overline{D}^{c}_\sigma(x,x')=
\int\! \frac{d^3 \tilde k}{(2\pi)^4}\int_{\cal C} dk_z\,
e^{-i\tilde k(\tilde x-\tilde x')+ik_z(z+z')}R^L_\sigma\frac{1}{k^2+i\epsilon}\;.
\nonumber\\
\label{eq:photonprop}
\end{eqnarray}
The derivatives of the polarization vectors (\ref{eq:polvecs}) act on a
plane wave and its reflection. We obtain
\begin{eqnarray}
e^{\mu}_\sigma(\partial_x)\,
            e^{\nu *}_\sigma(\partial_{x'})\; g^{\sigma \sigma}\;
e^{-i\tilde k(\tilde x-\tilde x')+ik_z(z+z')}\nonumber\\
=g^{\mu\nu}_\sigma(k)\;
e^{-i\tilde k(\tilde x-\tilde x')+ik_z(z+z')}\nonumber
\end{eqnarray}
with
\begin{eqnarray}
 g^{\mu\,\nu}_{TE}(k) &=&g^{m\,n}-\frac{k_m k_n}{-k_\parallel^2}=
              -\left(\delta_{m\,n}-\frac{k_m k_n}{k_\parallel^2}\right), 
\nonumber\\
 g^{\mu\,\nu}_{TM}(k) &=&\left\{
           g^{mn}= \frac{k_m k_n k_z^2}{{\bf k}^2 k_\parallel^2},\ 
           g^{m3}= \frac{k_m k_z}{{\bf k}^2 },\right.\nonumber\\&&\left.
\hspace*{3mm}    g^{3m}= -\frac{k_z^* k_m}{{\bf k}^2 },\ 
           g^{33}= -\frac{k_\parallel^2}{{\bf k}^2 }    \right\}
\label{eq:gmunusigma} \\
    g^{\mu\nu}_C(k) &=& g^{00}\,, \nonumber \\ 
    g^{\mu\nu}_G(k) &=& \left\{g^{mn}=-\frac{k_m k_n}{{\bf k}^2},\ 
                  g^{33} = \frac{k_z^2}{{\bf k}^2 },\right. \nonumber\\ &&\left.
\hspace*{3mm}     g^{m3} = \frac{k_zk_m}{{\bf k}^2 }=-g^{3m}
                        \right\}\nonumber 
\end{eqnarray}
where $m,n=1,2$. For evanescent waves one has $k_z=-i\kappa$,
$k_z^2=-\kappa^2$, and ${\bf k}^2=k_\parallel^2-\kappa^2$.

Inserting the electron propagator (\ref{elepr}) and  
the photon propagator (\ref{eq:photonprop})  into the expression for the
self energy (\ref{self0}), we have to multiply several $\gamma$
matrices. Using
\begin{eqnarray}
 \gamma^\mu \gamma^\lambda \gamma^\nu &\!\!=&\!\!
          s^{\alpha \mu \lambda\nu} \gamma_\alpha  
         -i \epsilon^{\alpha \mu \lambda\nu} \gamma_5 \gamma_\alpha,
 \hspace*{3mm}  \epsilon_{0123}=+1,  \nonumber\\
  s^{\alpha \mu \lambda\nu}&\!\!=&\!\! g^{\alpha \mu}g^{\lambda \nu}     
                             + g^{\alpha \nu}g^{\lambda \mu} 
                             - g^{\alpha \lambda}g^{\nu \mu}\;,
\label{gamma}
\end{eqnarray}
we encounter the following $\gamma$ valued invariants
\begin{eqnarray}
I_1^\sigma &\!\!=&\!\! m g^{\mu \nu} g_{\mu \nu}^\sigma\,,\ \ \ \ 
I_{15}^\sigma =\frac{m}{2} [\gamma^{\mu},\gamma^{\nu}] g_{\mu \nu}^\sigma\,,\nonumber\\ 
I_{25}^\sigma&\!\!=&\!\!\epsilon^{\alpha \mu \lambda\nu} \gamma_5 \gamma_\alpha
               g_{\mu \nu}^\sigma q_\lambda\, ,\nonumber \\
I_2^\sigma &\!\!=&\!\!  s^{\alpha \mu \lambda\nu} \gamma_\alpha g_{\mu \nu}^\sigma
                q_\lambda
            =  g^{\mu\nu}_\sigma \gamma_\mu q_\nu 
              +g^{\nu\mu}_\sigma \gamma_\mu q_\nu  
              - (\gamma q)I_1^\sigma/m\,,\nonumber 
\end{eqnarray}
with $q$ being the Fourier variable in the electron propagator (\ref{elepr})
and $g_{\mu \nu}^\sigma$ having the $k$ dependence as in (\ref{eq:gmunusigma}).
In terms of those the distance-dependent part of the self energy is
\begin{eqnarray}
\label{self2}
\Sigma =  \sum_\sigma \Sigma^\sigma\ , \hspace*{5mm}
\Sigma^\sigma=\left[\Sigma^\sigma_1 + \Sigma^\sigma_{15} + 
\Sigma^\sigma_2 + \Sigma^\sigma_{25} \right]
\end{eqnarray}
with
\begin{eqnarray}
    \Sigma^\sigma_\theta(x,x')
            &\!\!=&\!\! -ie^2\!  \int \frac{d^4 q}{(2\pi)^4}
		 \int \frac{d^3\tilde k}{(2\pi)^4}\int_C dk_z
                 \frac{e^{-iq(x-x')}} {q^2 - m^2 + i\epsilon}
\nonumber \\ && \times \            
I_\theta^\sigma(q, k)
\frac{e^{-i\tilde k (\tilde x-\tilde x')+ i k_z (z+z')} }{k^2  + i\epsilon} 
                  R^L_\sigma\;. \label{self4}
\end{eqnarray}
In the same way as for the total self energy in Eq.~(\ref{h6}) we perform a
Fourier transformation for the components in the sum (\ref{self2}), which
again retain a dependence on $z+z'$ due to the broken translation invariance
in the $z$ direction,
\begin{equation}
    \Sigma^\sigma_\theta(x,x')
            =   \int \frac{d^4p}{(2\pi)^4}\;
                 e^{-ip(x-x')}\,\Sigma^\sigma_\theta(p,z+z')\;. \nonumber
\end{equation}
Making the variable replacements $\tilde q=\tilde p - \tilde k$, ${q}_z =
p_z$ in Eq.~(\ref{self4}) we obtain
\begin{widetext}
\begin{equation}
\label{self5}
\Sigma^\sigma_\theta(p,z+z')= -ie^2\int \frac{d^3\tilde k}{(2\pi)^4}
\int_C dk_z\; \frac{1}{(\tilde p - \tilde k)^2 -p_z^2-m^2 +i\epsilon} 
\;I_\theta^\sigma(\tilde p-\tilde k, p_z, k)\;
\frac{e^{ i k_z(z+ z') } }{\tilde k^2  - k_z^2 + i\epsilon}
R^L_\sigma \;. 
\end{equation}
This expression looks much simpler than it is to evaluate. The loss of
translation invariance perpendicular to the surface of dielectric is one
source of complications, and the interference of incident and reflected
waves is another. In order to evaluate the self energy components
(\ref{self5}) we need the explicit expressions for the 
invariants $I^\sigma_\theta$, which depend on the mode $\sigma$. For the two
physical modes we find
\begin{eqnarray}
I_1^{TE} &\!\!=&\!\! m\,,\ \ I_{15}^{TE}=0\,, \ \ I_{25}^{TE}= 0\,,\nonumber\\
I_2^{TE} &\!\!=&\!\!-2\left[\boldsymbol\gamma_\parallel\cdot ({\bf p}_\parallel
-{\bf k}_\parallel) - \boldsymbol\gamma_\parallel\cdot {\bf k}_\parallel
\frac{{\bf k}_\parallel\cdot ({\bf p}_\parallel-{\bf k}_\parallel)}{k_\parallel^2} 
\right]+ \boldsymbol\gamma_\parallel\cdot ({\bf p}_\parallel-{\bf k}_\parallel)
+ \gamma_3 p_z  - \gamma_0(p_0-k_0)\,,            
\label{ITE}
\end{eqnarray}
\begin{eqnarray}
 I_1^{TM} &\!\!=&\!\! m\left(1 - 2\frac{k_z^2}{{\bf k}^2 }\right)\,,\ \  
 I_{15}^{TM}= -m\frac{1}{{\bf k}^2}\,[\gamma^3, \gamma^n]\,k_z k^n\,, \ \ 
 I_{25}^{TM}= 2\epsilon^{\alpha m \lambda 3}\gamma_5 \gamma_\alpha
                        (p_\lambda-k_\lambda) \frac{k_{m}k_z}{{\bf k}^2},
                      \nonumber \\
 I_2^{TM} &\!\!=&\!\! 
     2\left\{ \frac{(\boldsymbol\gamma_\parallel\cdot{\bf k}_\parallel) 
\left[{\bf k}_\parallel\cdot ({\bf p}_\parallel-{\bf k}_\parallel)\right]}
     {k_\parallel^2{\bf k}^2  } k_z^2
     -p_z \gamma_3 \frac{k_\parallel^2}{{\bf k}^2 } \right\}
 +\left[\boldsymbol\gamma_\parallel\cdot({\bf p}_\parallel-{\bf k}_\parallel)
+\gamma_3 p_z - \gamma_0 (p_0-k_0) \right]
     \left(1-2\frac{k_z^2}{{\bf k}^2 }\right).
\label{ITM}
\end{eqnarray}
\end{widetext}

\section{Asymptotic analysis of the self-energy\label{Sec:asympt}}
\subsection{General approach and approximations\label{subsec:intro}}
Our aim is to determine the energy shift of an electron that is localized in
$z$-direction. The shift will depend on the distance $a$ of the electron
from the surface of the dielectric, and without localization the notion of
this distance would not make sense. Physically the localization could be
realized be sending a tightly focussed beam parallel to the surface or by
confining the electron by means of magnetic and/or electric fields.  So, we
will in fact not be working directly with momentum eigenstates (\ref{h3})
but use them to form with wave packets that peak at $z=a$ and whose average
momentum in $z$ direction is $p_z$. The wave packet may move as a whole,
which is why we have not approximated ${\bf z}-{\bf z}'$ in
Eq.~(\ref{eq:dEdiscrete}); but we shall assume the electromagnetic field to
be the same across the packet, which corresponds to the dipole approximation
in atomic physics, and which is why we have set $z_3+z_3'\simeq -2a$ in
Eq.~(\ref{eq:dE}).

The extent of the wave packet must be small compared with the distance $a$
from the surface, but otherwise the details of the wave packet are not
relevant.  This implies $a m\gg 1$, i.e. that the distance $a$ must be very
much larger than the Compton wavelength $\lambdabar_C=\hbar/(mc)$. Thus, we
shall aim for an expansion in $1/(a p_0)$.

In order to proceed with the calculation of the self energy (\ref{self5}),
we want to perform a Wick rotation $k_0=ik_4$. By design, the poles of the
photon propagator lie in the right position for this. However, the poles of
the term that originates from the electron propagator may interfere; they
lie at \mbox{$k_0 = p_0 \pm [ \sqrt{({\bf p}_\parallel-{\bf
k}_\parallel)^2+p_z^2+m^2} - i\epsilon ]$}. There is no problem if they come
to lie in the 2nd and 4th quadrant of the complex $k_0$ plane, but for
\mbox{$2{\bf p}_\parallel\cdot{\bf k}_\parallel >{\bf k}_\parallel^2$}, one of the
poles lies in the 1st rather than the 2nd quadrant, if we take $p$ to be
on-shell, and a pole in the 1st quadrant interferes with the Wick
rotation. There are several ways of dealing with this problem. One could
work with a strongly deformed integration path and then carry along the
separate contribution from the pole; or one could go off-shell to move the
pole out of the 1st quadrant and then do an analytic continuation to a
result for on-shell $p$; or one could avoid the problem altogether by
approximating $p_\parallel\approx 0$ in the denominator of (\ref{self5}). We have
decided on the last approach because it is straightforward and we are not
interested in ultrarelativistic motion of the electron.

Thus we set $p_\parallel= 0$ in the denominator of (\ref{self5}) but leave  $p_\parallel$
untouched elsewhere, i.e.~retain it in $I^\sigma_\theta$ (\ref{ITE}) and
(\ref{ITM}). Then we can perform the Wick rotation $k_0=ik_4$ without
problems and obtain
\begin{eqnarray}
\Sigma^\sigma_\theta(p,z+z')= e^2\int \frac{d^2k_\parallel}{(2\pi)^4} 
\int_{-\infty}^{\infty} dk_4
\frac{1}{k_4^2+2ip_0k_4+{\bf k}_\parallel^2}\nonumber\\ 
\times\int_C dk_z\;I_\theta^\sigma
\frac{e^{ i k_z(z+ z') } }{k_4^2+{\bf k}_\parallel^2+ k_z^2}
R^L_\sigma \;. \nonumber
\end{eqnarray}
The integration over the three dimensional (Euclidean) space $({\bf
k}_\parallel,k_4)$ can be carried out in spherical polar coordinates by
defining $k_4=\rho\cos\vartheta, k_x=\rho\sin\vartheta\cos\varphi,
k_y=\rho\sin\vartheta\sin\varphi$. We find
\begin{eqnarray}
\Sigma^\sigma_\theta(p,z+z')= \frac{e^2}{(2\pi)^3}\int_0^\infty d\rho
\int_{-1}^{1} d(\cos\vartheta)
\frac{\rho^2}{\rho^2+2ip_0\rho\cos\vartheta}\nonumber\\ 
\times\int_C dk_z\;\frac{e^{ i k_z(z+ z') } }{\rho^2+ k_z^2}
R^L_\sigma\ \frac{1}{2\pi}\int_0^{2\pi}\!\!d\varphi\;I_\theta^\sigma\;. 
\hspace*{5mm}\label{eq:sigmaTETM}
\end{eqnarray}
The only $\varphi$ dependence is in the invariants $I_\theta^\sigma$;
carrying out the integration and using the fact that $p$ is on-shell, one gets
\begin{eqnarray}
\frac{1}{2\pi}\int_0^{2\pi}d\varphi\;\left(I_1^{TE}+I_2^{TE}\right)&\!\!=&\!\!\gamma_0 
i\rho\cos\vartheta - \boldsymbol\gamma_\parallel\cdot{\bf p}_\parallel\;,\nonumber\\
\frac{1}{2\pi}\int_0^{2\pi}d\varphi\;\left(I_1^{TM}+I_2^{TM}\right)&\!\!=&\!\!-\gamma_0 
i\rho\cos\vartheta + \boldsymbol\gamma_\parallel\cdot{\bf p}_\parallel\nonumber\\
&&\hspace{-15mm}+\frac{{\bf k}_\parallel^2}{{\bf k}^2}\left( 2\gamma_0i\rho\cos\vartheta
-2\gamma_3p_z-\boldsymbol\gamma_\parallel\cdot{\bf p}_\parallel\right)\,,\nonumber\\
\int_0^{2\pi}d\varphi\;I_{15}^{TM}=0\;,&&\int_0^{2\pi}d\varphi\;I_{25}^{TM}=0\;.
\nonumber
\end{eqnarray}
The next step in the evaluation of Eq.~(\ref{eq:sigmaTETM}) is to carry out
the integration over $k_z$ by means of contour integration. The remaining
two-dimensional integral over $\rho$ and $\cos\vartheta$ can then be
calculated asymptotically for $|z+z'|$ very much larger than the Compton
wavelength. Since the technical details differ between the $TE$ and $TM$
polarizations, we consider their contributions one after the other.
While the calculation to follow is perfectly general for all values of $z$
and $z'$, provided $|z+z'|p_0\gg 1$, we now simplify the notation and set
$z+z'=-2a$, as this is the value at which we need to evaluate the self
energy in Eq.~(\ref{eq:dE}) for the radiative shift.

\subsection{TE contributions to the self energy}
For the TE polarization the integrand of (\ref{eq:sigmaTETM}) has only one
pole in the lower $k_z$ plane, and that is at $k_z=-i\rho= -i({\bf
k}_\parallel^2+k_4^2)^{1/2}$ (cf.~Fig.~\ref{fig:contour}). The $k_z$
integration can thus easily be carried out by deforming the contour ${\cal
C}$ and evaluating the residue at $-i\rho$. We emphasize that when
evaluating the reflection coefficient $R^L_{TE}$, Eq.~(\ref{leftph}), at
this point, one must take great care that the branch cut of the square root
in $k_z^d$ is indeed taken to run as shown in
Fig.~\ref{fig:contour}. Renaming $\cos\vartheta=t$, we can write the result
of the contour integration as
\begin{eqnarray}
\Sigma^{TE}&\!\!=&\!\!\frac{e^2}{8\pi^2}\int_0^\infty\!\! d\rho \int_{-1}^1\! dt\;
\frac{\gamma_0 i\rho t - \boldsymbol\gamma_\parallel\cdot{\bf p}_\parallel}{
\rho+2ip_0t} R^L_{TE}(t)\;e^{-2a\rho}\nonumber\\
&\!\!=&\!\!\frac{e^2}{4\pi^2}\int_0^\infty\!\! d\rho \int_0^1\! dt\;\rho\;
\frac{2\gamma_0 p_0 t^2 - \boldsymbol\gamma_\parallel\cdot{\bf p}_\parallel}{
\rho^2+4p_0^2t^2} R^L_{TE}(t)\;e^{-2a\rho}\nonumber
\end{eqnarray}
with
\begin{equation}
R^L_{TE}(t)=\frac{1-\sqrt{(n^2-1)t^2+1}}{1+\sqrt{(n^2-1)t^2+1}}\;.
\end{equation}
Next we scale the integration variable $\rho=2p_0t\xi$. In terms of the new
variable $\xi$ the integral reads
\begin{eqnarray}
\Sigma^{TE} = \frac{e^2}{4\pi^2}\int_0^\infty d\xi \int_0^1 dt\;\xi\;
\frac{2\gamma_0 p_0 t^2 - \boldsymbol\gamma_\parallel\cdot{\bf p}_\parallel}{
\xi^2+1}\nonumber\\ \times\; R^L_{TE}(t)\;e^{-4p_0at\xi}\;.
\label{eq:TEint}
\end{eqnarray}
This integral can be evaluated asymptotically for large values of $p_0a$. A
standard method of obtaining an asymptotic expansion for integrals with an
exponentially damped integrand is repeated integration by parts. However, in
two-dimensional integrals like the one above, this method generally fails
because integration by parts in one variable generates inverse powers of the
other variable and the resulting integral diverges at the lower limit. For a
general discussion of this problem and its remedy, we refer the reader to
Ref.~\cite{siklos}. Here we observe that $R^L_{TE}(t)$ and hence the
integrand of (\ref{eq:TEint}) actually behave as $O(t^2)$ for $t\rightarrow
0$. Thus we can integrate by parts in the $\xi$ integral twice without
jeopardizing the convergence of the $t$ integral. In this way we find to
leading order in $1/(p_0a)$
\begin{equation}
\Sigma^{TE}\approx \frac{e^2}{4\pi^2} \frac{1}{(4p_0a)^2} \int_0^1 dt\;
\frac{2\gamma_0 p_0 t^2 - \boldsymbol\gamma_\parallel\cdot{\bf
p}_\parallel}{t^2}R^L_{TE}(t)\;.
\label{eq:finalTE}
\end{equation}
The $t$ integral in this expression is elementary.

\subsection{TM contributions to the self energy}

The TM polarization is more difficult to deal with, since the invariants
$I_\theta^{TM}$ introduce a factor $1/{\bf k}^2$ into the integrand of
(\ref{eq:sigmaTETM}), which leads to an additional pole in the lower $k_z$
plane at $k_z=-i\rho\sin\vartheta=-ik_\parallel$ (cf.~Fig.~\ref{fig:contour}).
Thus, closing the contour in the lower $k_z$ plane picks up two residues, one
at $-i\rho$ and one at $-i\rho\sin\vartheta$. The result is
\begin{widetext}
\begin{eqnarray}
\Sigma^{TM}&\!\!=&\!\!\frac{e^2}{8\pi^2}\int_0^\infty\!\! d\rho \int_{-1}^1\! dt\;
\frac{1}{\rho+2ip_0t}\left\{\left[ 
\boldsymbol\gamma_\parallel\cdot{\bf p}_\parallel - \gamma_0 i\rho t 
+\frac{1-t^2}{t^2} \left(\boldsymbol\gamma_\parallel\cdot{\bf p}_\parallel 
- 2\gamma_0 i\rho t + 2\gamma_3 p_z\right)\right]
R^L_{TM}(t)\; e^{-2a\rho}\right.\nonumber\\
&&\hspace*{60mm}\left.-\frac{\sqrt{1-t^2}}{t^2}
\left(\boldsymbol\gamma_\parallel\cdot{\bf p}_\parallel 
- 2\gamma_0 i\rho t + 2\gamma_3 p_z\right)
R^L_{TM}(0)\; e^{-2a\rho\sqrt{1-t^2}}\right\}\nonumber\\
&\!\!=&\!\!\frac{e^2}{4\pi^2}\int_0^\infty\!\! d\rho \int_0^1\! dt\;
\frac{\rho}{\rho^2+4p_0^2t^2}\left\{\left[
\frac{1}{t^2}\boldsymbol\gamma_\parallel\cdot{\bf p}_\parallel -2\gamma_0p_0
\left(2-t^2\right)+\frac{1-t^2}{t^2} 2\gamma_3 p_z \right]
R^L_{TM}(t)\; e^{-2a\rho}\right.\nonumber\\
&&\hspace*{60mm}\left.-\frac{\sqrt{1-t^2}}{t^2}
\left(\boldsymbol\gamma_\parallel\cdot{\bf p}_\parallel 
- 4\gamma_0p_0t^2 + 2\gamma_3 p_z\right)
R^L_{TM}(0)\; e^{-2a\rho\sqrt{1-t^2}}\right\}
\label{eq:TMint}
\end{eqnarray}
where we have again renamed $\cos\vartheta=t$ and abbreviated
\[
R^L_{TM}(t)=\frac{n^2-\sqrt{(n^2-1)t^2+1}}{n^2+\sqrt{(n^2-1)t^2+1}}\;.
\]
As before, we are interested in an asymptotic result for $\Sigma^{TM}$ for
large values of $p_0a$. To be able to do asymptotic analysis one needs to
separate the terms with different arguments in the exponential. However,
doing this simple-mindedly leads to two divergent integrals because their
integrands each behave as $O(t^{-2})$ for $t\rightarrow 0$. That is why we
add and subtract the same term and subdivide the integral as follows,
\begin{equation}
\Sigma^{TM} = \Sigma^{TM}_A + \Sigma^{TM}_B + \Sigma^{TM}_C +
\Sigma^{TM}_D\;,
\label{eq:sigmasum}
\end{equation}
with
\begin{eqnarray}
\Sigma^{TM}_A&\!\!=&\!\!\frac{e^2}{4\pi^2}\int_0^\infty\!\! d\xi \int_0^1\! dt\;
\frac{\xi}{\xi^2+1}\left[ 2\gamma_0p_0 \left(t^2-2\right) -2\gamma_3 p_z\right]
R^L_{TM}(t)\; e^{-4p_0at\xi} \nonumber\\
\Sigma^{TM}_B&\!\!=&\!\!\frac{e^2}{4\pi^2}\left(2\gamma_3 p_z 
+\boldsymbol\gamma_\parallel\cdot{\bf p}_\parallel \right)
\int_0^\infty\!\! d\xi \int_0^1\! dt\;\frac{\xi}{\xi^2+1}\frac{1}{t^2}
\left[ R^L_{TM}(t)-R^L_{TM}(0)\right]\; e^{-4p_0at\xi} \nonumber\\
\Sigma^{TM}_C&\!\!=&\!\!\frac{e^2}{4\pi^2}\left(2\gamma_3 p_z 
+\boldsymbol\gamma_\parallel\cdot{\bf p}_\parallel \right) R^L_{TM}(0)
\int_0^\infty\!\! d\xi \int_0^1\!dt\;\frac{\xi}{\xi^2+1}
\frac{1}{t^2} \left(e^{-4p_0at\xi} 
- \sqrt{1-t^2}e^{-4p_0at\sqrt{1-t^2}\xi} \right)\nonumber\\
\Sigma^{TM}_D&\!\!=&\!\!\frac{e^2}{4\pi^2} 4\gamma_0 p_0  R^L_{TM}(0)
\int_0^\infty\!\!d\xi \int_0^1\!
dt\;\frac{\xi}{\xi^2+1}\sqrt{1-t^2}\;e^{-4p_0at\sqrt{1-t^2}\xi}
\nonumber
\end{eqnarray}
where we have again rescaled $\rho=2p_0t\xi$. 

We do the asymptotic analysis of these integrals one by one, starting with
$\Sigma^{TM}_A$. In order to get an asymptotic expansion for large $p_0a$,
one would try to integrate by parts. However, The integrand of
$\Sigma^{TM}_A$ behaves as $O(1)$ for $t\rightarrow 0$, and thus the factor
$1/t$ that one gets through integrating by parts in the $\xi$ integral would
destroy the convergence at $t=0$. Adapting the general method of obtaining
an asymptotic expansion of such two-dimensional integrals \cite{siklos}, we
add and subtract the problematic point at $t=0$ and write
\begin{eqnarray}
\Sigma^{TM}_A&\!\!=&\!\!-\frac{e^2}{4\pi^2}\int_0^\infty\!\! d\xi \int_0^1\! dt\;
\frac{\xi}{\xi^2+1}\left( 4\gamma_0p_0 +2\gamma_3 p_z\right) R^L_{TM}(0)\; 
e^{-4p_0at\xi} \nonumber\\
&&+\frac{e^2}{4\pi^2}\int_0^\infty\!\! d\xi \int_0^1\! dt\;
\frac{\xi}{\xi^2+1}\left\{\left[ 2\gamma_0p_0 \left(t^2-2\right) -2\gamma_3
p_z\right]R^L_{TM}(t)
+\left(4\gamma_0p_0 +2\gamma_3 p_z\right) R^L_{TM}(0)\right\} e^{-4p_0at\xi} \nonumber
\end{eqnarray}
The first of the integrals is easy to calculate; the $t$ integral is
immediate, and the remaining integral over $\xi$ is a well known combination
of sine and cosine integrals \cite{as}. The integrand of the
second integral now behaves as $O(t^2)$ for $t\rightarrow 0$, and we can
thus integrate by parts twice without getting convergence problems at
$t=0$. In this way we find to order $1/(p_0a)^2)$
\begin{eqnarray}
\Sigma^{TM}_A \approx -\frac{e^2}{8\pi^2p_0a} \left( 2\gamma_0p_0 +\gamma_3 p_z\right)
R^L_{TM}(0) \left(\frac{\pi}{2} - \frac{1}{4p_0 a}\right)+ \frac{e^2}{32\pi^2p_0^2a^2}
\left\{ \gamma_0 p_0 \int_0^1 \frac{dt}{t^2} \left[ \left(t^2-2\right)\;R^L_{TM}(t)
+ 2\;R^L_{TM}(0) \right] \right.\nonumber\\
\left.- \gamma_3 p_z \int_0^1 \frac{dt}{t^2} \left[ R^L_{TM}(t)
- R^L_{TM}(0) \right] \right\}\hspace*{5mm}\label{eq:sigmaA}
\end{eqnarray}
\end{widetext}
for which we have also made use of the known asymptotics of the sine and
cosine integrals \cite{as}.

The asymptotics of $\Sigma^{TM}_B$ can be calculated similarly by adding and
subtracting the next term in the Taylor expansion of $R^L_{TM}(t)$, i.e. by replacing
\begin{eqnarray}
&&\hspace*{-5mm}
\frac{1}{t^2}\left[ R^L_{TM}(t)-R^L_{TM}(0)\right] \nonumber\\
&&\hspace*{-5mm} = R^{L}_{TM}{''}(0)/2
+\frac{1}{t^2}\left[R^L_{TM}(t)-R^L_{TM}(0)- t^2 R^{L}_{TM}{''}(0)/2 \right].\nonumber
\end{eqnarray}
The first term can then be integrated exactly, and the rest behaves as
$O(t^2)$ for $t\rightarrow 0$ and can thus integrated by parts with respect
to $\xi$ twice. The result to order $1/(p_0a)^2)$ is
\begin{widetext}
\begin{equation}
\Sigma^{TM}_B \approx -\frac{e^2}{4\pi^2p_0a} \left( 2\gamma_3 p_z + 
\boldsymbol\gamma_\parallel\cdot{\bf p}_\parallel\right) \frac{1}{4p_0 a}
\left\{ \frac{1}{2}R^{L}_{TM}{''}(0) \left(\frac{\pi}{2}-\frac{1}{4p_0 a}\right)
+\frac{1}{4p_0 a} \int_0^1 \frac{dt}{t^2} \left[
\frac{R^L_{TM}(t)-R^L_{TM}(0)}{t^2}-\frac{1}{2}R^{L}_{TM}{''}(0)\right]\right\}
\label{eq:sigmaB}
\end{equation}

Next we turn our attention to the asymptotic evaluation of
$\Sigma^{TM}_C$. We cannot separate the two summands in the integrand
because otherwise the $t$ integral does not converge. Thus, to manipulate
just one part, we must set the lower limit of the $t$ integral to some small
positive $\varepsilon$ and take the limit $\varepsilon\rightarrow 0$ only
once we have combined all parts again. We write
$$
\int_\varepsilon^1 dt \frac{\sqrt{1-t^2}}{t^2} e^{-4p_0at\sqrt{1-t^2}\xi}
= \int_\varepsilon^{1/\sqrt{2}} dt \frac{\sqrt{1-t^2}}{t^2}e^{-4p_0at\sqrt{1-t^2}\xi}
+ \int_{1/\sqrt{2}}^1 dt \frac{\sqrt{1-t^2}}{t^2} e^{-4p_0at\sqrt{1-t^2}\xi}\;.
$$
In the integral that runs from $1/\sqrt{2}$ to 1 we make a change of
variable from $t$ to $s=\sqrt{1-t^2}$. Then renaming $s$ into $t$ again and
ignoring terms that vanish in the limit $\varepsilon\rightarrow 0$, we find
$$
\int_\varepsilon^1 dt \frac{\sqrt{1-t^2}}{t^2} e^{-4p_0at\sqrt{1-t^2}\xi}
= \int_\varepsilon^{1/\sqrt{2}} dt \left[\frac{\sqrt{1-t^2}}{t^2}+ 
\frac{t^2}{\left(1-t^2\right)^{3/2}}\right] e^{-4p_0at\sqrt{1-t^2}\xi}\;.
$$
Now we substitute $v=t\sqrt{1-t^2}$ and obtain
$$
\int_\varepsilon^1 dt \frac{\sqrt{1-t^2}}{t^2} e^{-4p_0at\sqrt{1-t^2}\xi}
=\int_\varepsilon^{1/2} dv \frac{1-2v^2}{v^2\sqrt{1-4v^2}} e^{-4p_0av\xi}\;.
$$
Renaming the integration variable $v$ into $t$ again, we can use this
identity to write $\Sigma^{TM}_C$ as
$$
\Sigma^{TM}_C=\frac{e^2}{4\pi^2}\left(2\gamma_3 p_z 
+\boldsymbol\gamma_\parallel\cdot{\bf p}_\parallel \right)R^L_{TM}(0)
\int_0^\infty\!\! d\xi \frac{\xi}{\xi^2+1}\left[\int_0^{1/2}\!dt
\frac{1}{t^2} \left(1-\frac{1-2t^2}{\sqrt{1-4t^2}}\right)e^{-4p_0at\xi}
+ \int_{1/2}^1\!dt\frac{1}{t^2}e^{-4p_0at\xi}\right]\;.
$$
\end{widetext}
Integrating by parts in the $\xi$ integral twice, we obtain to order $1/(p_0a)^2)$ 
\begin{equation}
\Sigma^{TM}_C \approx -\frac{e^2}{4\pi^2}\left(2\gamma_3 p_z 
+\boldsymbol\gamma_\parallel\cdot{\bf p}_\parallel \right)R^L_{TM}(0) 
\frac{1}{48p_0^2a^2}\;.\label{eq:sigmaC}
\end{equation}

For the asymptotic expansion of $\Sigma^{TM}_D$ we apply much the same
tricks as above: we split the $t$ integration at $1/\sqrt{2}$ and in the
integral from $1/\sqrt{2}$ to 1 we substitute $s=\sqrt{1-t^2}$. Then we rename $s$
back to $t$, combine the two integrals again, and substitute
$v=t\sqrt{1-t^2}$ to obtain
\begin{eqnarray}
\Sigma^{TM}_D&\!\!=&\!\!\frac{e^2}{\pi^2} \gamma_0 p_0  R^L_{TM}(0)
\int_0^\infty\!\!d\xi \int_0^{1/2}\!
dv\;\frac{\xi}{\xi^2+1}\frac{1}{\sqrt{1-4v^2}}\nonumber\\
&&\hspace*{52mm}\times\ e^{-4p_0av\xi}\nonumber\\
&\!\!=&\!\!\frac{e^2}{\pi^2} \gamma_0 p_0  R^L_{TM}(0) \left[ 
\int_0^\infty\!\!d\xi \int_0^{1/2}\!dv\;\frac{\xi}{\xi^2+1} e^{-4p_0av\xi}
\right.\nonumber\\
&&\!\!+\left.\int_0^\infty\!\!d\xi \int_0^{1/2}\!
dv\;\frac{\xi}{\xi^2+1}\left(\frac{1}{\sqrt{1-4v^2}}-1\right) 
e^{-4p_0av\xi}\right].\nonumber
\end{eqnarray}
The first of those integrals can be solved in terms of known special
functions \cite{as}, and in the second integral we can integrate by parts
twice with respect to $\xi$. Thus we obtain for $\Sigma^{TM}_D$ to order 
$1/(p_0a)^2)$
\begin{equation}
\Sigma^{TM}_D \approx \frac{e^2}{8\pi a} \gamma_0 R^L_{TM}(0)\;.\label{eq:sigmaD}
\end{equation}

According to (\ref{eq:sigmasum}) we combine the results 
(\ref{eq:sigmaA}--\ref{eq:sigmaD}) to obtain the leading-order self-energy contribution
from the TM modes. It turns out to be one order larger than that from the TE
modes.
\begin{widetext}
\begin{eqnarray}
\Sigma^{TM} &\!\!\approx&\!\! - \frac{e^2}{32\pi p_0a} \left[ 
\boldsymbol\gamma_\parallel\cdot{\bf p}_\parallel 
\frac{n^2\left(n^2-1\right)}{\left(n^2+1\right)^2}
+ 2\gamma_3 p_z \frac{2n^4-n^2-1}{\left(n^2+1\right)^2} +
O\left(1/p_0a\right) \right]\;.
\nonumber
\end{eqnarray}
Thus, to leading order the radiative self energy is due to just the TM modes.
\begin{eqnarray}
\Sigma(p,-2a) &\!\!\approx&\!\! - \frac{e^2}{32\pi p_0a} \left[ 
\boldsymbol\gamma_\parallel\cdot{\bf p}_\parallel 
\frac{n^2\left(n^2-1\right)}{\left(n^2+1\right)^2} 
\label{eq:leading}
+ 2\gamma_3 p_z \frac{2n^4-n^2-1}{\left(n^2+1\right)^2}\right]
+\Sigma^{next}
\end{eqnarray}
The next-to-leading term $\Sigma^{next}$ is easily determined from the
results (\ref{eq:finalTE}) and (\ref{eq:sigmaA}--\ref{eq:sigmaD}). While all the
$t$ integrals are elementary, it is convenient to use formula manipulation
software like Maple to evaluate and combine them. The final result for the
next-to-leading order contribution to the radiative self-energy is
\begin{eqnarray}
\Sigma^{next} &\!\!=&\!\! - \frac{e^2}{32\pi^2 (p_0a)^2} \left\{
2\gamma_0 p_0\left[\frac{n^2+1}{\sqrt{n^2-1}}\ln\left(n+\sqrt{n^2-1}\right)
-\frac{n^3+4n^2+n+2}{\left(n^2+1\right)(n+1)}-\frac{2n^4}{\left(n^2+1\right)^{3/2}}
\mbox{arctanh}\frac{n-1}{\sqrt{n^2+1}}\right]\right.\nonumber\\&&\hspace*{23mm}
+\boldsymbol\gamma_\parallel\cdot{\bf p}_\parallel \left[ 
\frac{n^2(2n^4+3n^2-3n+1)}{3(n+1)(n^2+1)^2}-\frac{2n^4}{(n^2-1)\left(n^2+1\right)^{5/2}}
\mbox{arctanh}\frac{n-1}{\sqrt{n^2+1}}\right]\nonumber\\&&\hspace*{23mm}\left.
+\gamma_3 p_z \left[ \frac{2n^6+n^5-3n^4-4n^3-4n^2+n+1}{3(n+1)(n^2+1)^2}
+\frac{4n^6}{(n^2-1)\left(n^2+1\right)^{5/2}}
\mbox{arctanh}\frac{n-1}{\sqrt{n^2+1}}\right]\right\}.\nonumber
\end{eqnarray}
\end{widetext}

\subsection{Self energy in the limit $n\rightarrow\infty$}\label{sec:ninf}
In the limit of perfect reflectivity $n\rightarrow\infty$ the calculation of
the self energy simplifies considerably. All reflection coefficients go to
either +1 or -1 (cf.~Eq.~(\ref{inft})), and the photon propagator takes on a
much simpler form with the $k_3$ integration running straight along the real
axis (cf.~Eq.~(\ref{intfw})). The calculation of the self energy can then
proceed in exactly the same way as explained in section \ref{subsec:intro}
above. It starts to differ only with the asymptotic analysis of
Eqs.~(\ref{eq:TEint}) and (\ref{eq:TMint}). For TE the asymptotic
analysis of (\ref{eq:TEint}) relied on the fact that $R_{TE}^{L}(t)$ behaves
as $O(t^2)$ for $t\rightarrow 0$, but in the limit $n\rightarrow\infty$ we
have $R_{TE}^{L}=-1$ for all $t$, which leads to a very different asymptotic
behaviour of the self energy $\Sigma^{TE}$. To leading order we find
\[
\Sigma^{TE}_{\mbox{\scriptsize\it perf}}\approx \frac{e^2}{32\pi p_0 a}
\boldsymbol\gamma_\parallel\cdot{\bf p}_\parallel\;.
\]
In the case of TM, something similar happens. The integrals $\Sigma_C^{TM}$,
$\Sigma_D^{TM}$, and to leading order in $1/p_0a$ also $\Sigma_A^{TM}$ give
the same with the limit $n\rightarrow\infty$ taken first as they do for
finite $n$ and with the limit $n\rightarrow\infty$ taken in the end result
of the asymptotic calculation. However, the integral $\Sigma_B^{TM}$ does
not even appear if the limit $n\rightarrow\infty$ is taken straightaway. For
finite $n$ its asymptotically leading term depends on the second derivative
$R_{TM}^{L}{''}(0)$, which is of course zero if the limit
$n\rightarrow\infty$ has been taken first and $R_{TM}^{L}$ is a constant. If
one takes $n\rightarrow\infty$ first then to leading order only
$\Sigma_A^{TM}$ and $\Sigma_D^{TM}$ contribute, and one obtains
\[
\Sigma^{TM}_{\mbox{\scriptsize\it perf}}\approx - \frac{e^2}{16\pi p_0 a} 
\gamma_3 p_z\;.
\]
In total the radiative part of the self energy with the perfect-reflector
limit taken first is
\[
\Sigma_{\mbox{\scriptsize\it perf}}(p,-2a)\approx \frac{e^2}{32\pi p_0 a} \left(
\boldsymbol\gamma_\parallel\cdot{\bf p}_\parallel -2 \gamma_3 p_z \right)\;,
\]
which clearly differs from the limit $n\rightarrow\infty$ of
Eq.~(\ref{eq:leading}).
So, mathematically not surprisingly, we find that the result for
the self energy differs depending on whether we perform the calculation for
finite $n$ and subsequently take the limit $n\rightarrow\infty$, or whether
we take the limit of perfect reflectivity $n\rightarrow\infty$ first and
evaluate the integrals then. The cause of this is simply the fact that the limits
$n\rightarrow\infty$ and $t\rightarrow 0$ in the Fresnel coefficients
$R_{\sigma}^{L}(t)$ do not commute. The point $t=0$ corresponds to
$\cos\vartheta=0$, i.e. to $k_0=0$. Thus, physically speaking, the limit of
perfect reflectivity is not interchangeable with the limit
$\omega\rightarrow 0$; in other words, long-wavelengths excitations get
treated very differently in and away from the limit of perfect reflectivity,
$n\rightarrow\infty$. Indications of this can also be seen in the
non-relativistic calculation of the energy level shift \cite{PRL}. 

The important lesson to be learnt from this observation is that models that
assume perfect reflectivity from the outset are bound to give the wrong
answer if long-wavelengths excitations play any role in the system under
investigation. Luckily, most of cavity QED is concerned with atoms and other
bound systems which have an inherent low-frequency cut-off (e.g.~the lowest
transition frequency $\omega_{ij}$ of an atom in state $|i\rangle$ to
dipole-allowed states $|j\rangle$). However, for unbound or partially bound
systems a perfect-reflector model is principally inadequate for describing
any physically realizable system, no matter how good the reflectivity of the
boundaries may be \cite{PRL}.

\subsection{Electrostatic contribution\label{Sec:elstatic}}

The evaluation of the electrostatic shift (\ref{h1c}) is straightforward.
The state in which the expectation value is being taken is a wave packet
that is localized at around ${\bf x}_a=(0,0,-a)$ and that has an average
momentum ${\bf p}$. Here the localization parallel to the surface of the
dielectric can of course be arbitrarily loose, as the system is translation
invariant parallel to the surface and hence the energy shift does not depend
on the transverse location of the particle. We choose to represent the
localized state by a Gaussian wave packet,
$$
|\varphi({\bf x}_a, {\bf p})\rangle = 
\frac{1}{\pi^{{3/4}}\sigma^{{3/2}}} \int d^3{\bf q}\;
e^{-\frac{({\bf q}-{\bf p})^2}{2\sigma^2}-i{\bf q}\cdot {\bf x}_a}\: 
b^\dagger_{{\bf q},i}|0\rangle\;.
$$
Taking the expectation value in Eq.~(\ref{h1c}) in this state and using the
canonical mode expansion (\ref{eq:spinorexpansion}) for the spinor operators, 
we obtain
\begin{widetext}
\begin{eqnarray}
\delta E _{\mbox{\scriptsize Coulomb}} = &&\hspace*{-3mm}
          \frac{e^2}{2\pi^{3/2}\sigma^3}\;\int d^3{\bf x} \int d^3{\bf x}'
\int d^3{\bf p}'\int \frac{d^3{\bf q}}{(2\pi)^3}
\int \frac{d^3{\bf q}'}{(2\pi)^3}\;
\frac{m^2}{p_0'\sqrt{q_0q_0'}} e^{i(q_0'-q_0)t-i({\bf q}-{\bf q}')\cdot{\bf x}_a 
+i({\bf q}-{\bf p}')\cdot{\bf x}'-i({\bf q}'-{\bf p}')\cdot{\bf x}}\nonumber\\
&&\hspace*{-3mm}\times e^{-\frac{({\bf q}-{\bf p})^2}{2\sigma^2}
-\frac{({\bf q}'-{\bf p})^2}{2\sigma^2} } \sum_{j=1,2}
\overline{u}^{(i)}({\bf q}')\;\gamma^0\; u^{(j)}({\bf p}')\;
\overline{u}^{(j)}({\bf p}')\;\gamma^0\; u^{(i)}({\bf q})\ 
\overline{G}_\Phi({\bf x},{\bf x}') 
\,. \nonumber 
\end{eqnarray}
\end{widetext}
Since the integrand as function of ${\bf x}$ and ${\bf x}'$ is peaked at the
location ${\bf x}_a$ of the wave packet, we can approximate the Green's
function $\overline{G}_\Phi({\bf x},{\bf x}')$ by 
$\overline{G}_\Phi({\bf x}_a,{\bf x}_a)$. Then the ${\bf x}$ and ${\bf x}'$ 
integrations are easy to carry out and give $\delta$ functions. The sum over 
polarizations $j$ can also be done because the Dirac eigenspinors satisfy
$$\sum_{j=1,2} u^{(j)}({\bf p}') \overline{u}^{(j)}({\bf p}') = \frac{1}{2m}
(\gamma^\mu p_\mu' +m)\;.
$$
Thus the above expression simplifies to
\begin{eqnarray}
\delta E _{\mbox{\scriptsize Coulomb}} \simeq \frac{e^2}{2\pi^{3/2}\sigma^3}\;
\overline{G}_\Phi({\bf x}_a,{\bf x}_a)\int d^3{\bf p}' \frac{m}{2{p'}_0^2}
e^{-\frac{({\bf p}'-{\bf p})^2}{\sigma^2}}
\nonumber\\ \times \;
\overline{u}^{(i)}({\bf p}') (\gamma^0 p_0' + 
\boldsymbol\gamma\cdot{\bf p}'+m)u^{(i)}({\bf p}')\;.\nonumber
\end{eqnarray}
The integrand of this expression peaks at ${\bf p}'={\bf p}$, so that we can
approximate ${\bf p}'$ by ${\bf p}$ everywhere except in the exponential and
carry out the integration. Taking into account that ${\bf p}$ is on shell,
we then find
$$
\delta E _{\mbox{\scriptsize Coulomb}} \simeq \frac{e^2}{2}\;
\overline{G}_\Phi({\bf x}_a,{\bf x}_a)\; \frac{m}{p_0}\; 
\overline{u}^{(i)}({\bf p}) \gamma^0 u^{(i)}({\bf p})\;.
$$

It remains the evaluation of the Green's function at the location of the
particle.  Since we are looking for the energy shift relative to a particle
in free space, we use the difference of the Coulomb Green's function
(\ref{greencphi}) and the free-space Coulomb Green's function, $$
\overline{G}_\Phi(x,x') = G_\Phi(x,x') - \int \frac{d^3k}{(2\pi)^3}\
e^{i({\bf x}-{\bf x}')\cdot {\bf k}}\; \frac{1}{{\bf k}^2}\;.$$ As ${\bf
x}_a$ is outside the dielectric, we have $$\overline{G}_{\Phi}({\bf
x}_a,{\bf x}_a)= - \frac{n^2-1}{n^2+1}
\frac{1}{4\pi^2} \int_0^{\infty}\!\! d k_\parallel \int_{-\infty}^\infty dk_z
\frac{k_\parallel\; e^{-2iak_z}}{k_\parallel^2+k_z^2}\,.
$$
For the $k_z$ integration we close the contour in the lower half-plane. The
integration over $k_\parallel$ is then trivial. The result is
$$
\overline{G}_{\Phi}({\bf x}_a,{\bf x}_a)= - \frac{n^2-1}{n^2+1} 
\frac{1}{8\pi a}\;.
$$
Thus the electrostatic energy shift is
\begin{equation}
\delta E _{\mbox{\scriptsize Coulomb}} \simeq - \frac{n^2-1}{n^2+1}
\;\frac{e^2}{16\pi a}\; \frac{m}{p_0}\; 
\overline{u}^{(i)}({\bf p}) \gamma^0 u^{(i)}({\bf p})\;.
\end{equation}
For a particle at rest this agrees with the classical energy shift of a
point particle in front of a dielectric half-space \cite{jackson}.

If one wishes, one can express this shift as being due to a Coulomb
self-energy function. If the shift is given by Eq.~(\ref{eq:dE}) then we can
write
\begin{equation}
\Sigma_{\mbox{\scriptsize Coulomb}}(p,-2a) \simeq - \frac{n^2-1}{n^2+1}
\;\frac{e^2}{16\pi a}\;\gamma^0\;.
\label{eq:SigmaES}
\end{equation}

\section{Summary and discussion of the results\label{Sec:final}} 
Combining our results for the radiative self-energy (\ref{eq:leading}) and for the
electrostatic self-energy (\ref{eq:SigmaES}), we obtain for the total self-energy
to leading order in $1/(p_0a)$
\begin{widetext}
\begin{equation}
\Sigma_{\mbox{\scriptsize total}}(p,-2a) \simeq - \frac{e^2}{32\pi p_0a} \left[ 
\boldsymbol\gamma_\parallel\cdot{\bf p}_\parallel 
\frac{n^2\left(n^2-1\right)}{\left(n^2+1\right)^2} 
+ 2\gamma_3 p_z \frac{2n^4-n^2-1}{\left(n^2+1\right)^2} +2\gamma^0p_0 
\frac{n^2-1}{n^2+1}\right]\;.
\label{eq:final}
\end{equation}
The total energy shift is easily determined from Eq.(\ref{eq:dE}). Since
spin-up and spin-down states are degenerate without the perturbation, the
right-hand side of Eq.(\ref{eq:dE}) is actually a matrix with the
energy shifts as eigenvalues. In general, the self-energy operator (\ref{eq:final}) is
not diagonal in the spin states $u^{(i)}({\bf p})$, as given in
Eq.~(\ref{eq:eigenspinors}), if spin up and down states are defined
along $z$, i.e. for
$$\phi^{(1)}=\left(\begin{array}{c}1\\0\end{array}\right),\ \ 
\phi^{(2)}=\left( \begin{array}{c}0\\1\end{array}\right) \;.$$
We find
\begin{equation}
\frac{m}{E}\overline{u}^{(1)}\Sigma_{\mbox{\scriptsize total}} u^{(1)}=
\frac{m}{E}\overline{u}^{(2)}\Sigma_{\mbox{\scriptsize total}} u^{(2)}=
-\frac{e^2}{32\pi a E^2} 
\left( 2\frac{2n^4-n^2-1}{(n^2+1)^2}\langle p_z^2\rangle 
+ \frac{n^2(n^2-1)}{(n^2+1)^2}\langle p_\parallel^2\rangle 
+ 2 \frac{n^2-1}{n^2+1} E^2 \right) \label{eq:diag}
\end{equation}
and
\begin{equation}
\frac{m}{E}\overline{u}^{(2)}\Sigma_{\mbox{\scriptsize total}} u^{(1)}=
\left(\frac{m}{E}\overline{u}^{(1)}\Sigma_{\mbox{\scriptsize total}} u^{(2)}
\right)^*=-\frac{e^2}{32\pi a E^2} 
\left(\frac{n^2(n^2-1)}{(n^2+1)^2} + 2 \frac{n^2-1}{n^2+1} \frac{E}{E+m}
\right) \langle (p_x + ip_y) p_z \rangle\;. \label{eq:nondiag}
\end{equation}
Thus the energy shift is
\begin{eqnarray}
\delta E = -\frac{e^2}{32\pi a E^2} 
\left[ 2\frac{2n^4-n^2-1}{(n^2+1)^2}\langle p_z^2\rangle 
+ \frac{n^2(n^2-1)}{(n^2+1)^2}\langle p_\parallel^2\rangle 
+ 2 \frac{n^2-1}{n^2+1} E^2\hspace*{20mm}\right.\nonumber\\ \left.
\pm \left( \frac{n^2(n^2-1)}{(n^2+1)^2} + 2 \frac{n^2-1}{n^2+1}
\frac{E}{E+m}
\right) \sqrt{\langle p_x p_z\rangle^2 +\langle p_y p_z\rangle^2}\right]\;.
\label{eq:finalshift}
\end{eqnarray}
\end{widetext}
For wave packets that are either stationary or whose motion preserves the
symmetry of the problem, the non-diagonal elements (\ref{eq:nondiag}) are
zero and the energy shift is given simply by (\ref{eq:diag}).

In the limit of perfect reflectivity $n\rightarrow\infty$ the calculation
yields a result for the total self-energy that differs from the limit
$n\rightarrow\infty$ of Eq.~(\ref{eq:final}),
\begin{equation}
\Sigma_{\mbox{\scriptsize total}}^{\mbox{\scriptsize\it perf}}(p,-2a) \simeq 
-\frac{e^2}{32\pi p_0a} \left( 
-\boldsymbol\gamma_\parallel\cdot{\bf p}_\parallel 
+ 2\gamma_3 p_z +2\gamma^0p_0 \right)\;.
\end{equation}
Accordingly, the energy shift differs from the limit
$n\rightarrow\infty$ of Eq.~(\ref{eq:finalshift}).
We have discussed the reasons for this discrepancy in Section \ref{sec:ninf}
and in Ref.~\cite{PRL}.

Quite apart from calculating the energy shift of an electron in front of an
imperfectly reflecting half-space, we have established the major building
blocks for QED in the presence of a dielectric half-space. Two alternative
formulations for the Feynman propagator of the electromagnetic field are
given in Eqs.~(\ref{fd2}) and (\ref{fd3}). The loss of translation
invariance perpendicular to the surface of the dielectric half-space is an
essential complication in loop calculations, but we have demonstrated how to
tackle this at one-loop level.

\begin{acknowledgments}
It is a pleasure to thank Gabriel Barton for many discussions. We are most
grateful to the Royal Society for financial support.
\end{acknowledgments}

\appendix
\section{Polarization vectors and normal modes\label{Sec:modes}}
In generalized Coulomb gauge the direction of the electromagnetic field can
be described by the following choice of polarization vectors:
\begin{eqnarray}
&&e^\mu_{TE}= -e_{\mu\,\,TE}= (-\Delta_{\parallel})^{-1/2}
(0,-i \partial_y,i\partial_x,0)\;,\nonumber\\
&&e^\mu_{TM}= -e_{\mu\,\,TM}= (\Delta \Delta_{\parallel})^{-1/2}
            (0,- \partial_x\partial_z,-\partial_y\partial_z,
             \Delta_{\parallel})\;,\nonumber \\
&&e^\mu_{G}= -e_{\mu\,\,G} = (-\Delta)^{-1/2}(0, -i \partial_x, -i\partial_y, 
           -i\partial_z)\;,               \nonumber  \\
&&e^\mu_{C}= e_{\mu\,\,C} =(1,0,0,0) \;. \label{eq:polvecs}
\end{eqnarray} 
Here $\Delta=\partial_x^2+\partial_y^2+\partial_z^2$ is the Laplacian in
three dimensions and $\Delta_\parallel=\partial_x^2+\partial_y^2$ is the one
in two dimensions parallel to the surface of the dielectric. The physical
polarizations are the transverse electric $e^\mu_{TE}$ and the transverse
magnetic $e^\mu_{TM}$, which have vanishing electric or magnetic,
respectively, field components perpendicular to the surface. The unphysical
polarizations are the longitudinal $e^\mu_{G}$ and the timelike $e^\mu_{C}$.

Constructing the normal modes is straightforward if one proceeds from a
plane incident wave which gets reflected and transmitted at the
surface. The incident wave is what it would be in free space (for
left-incident modes) or in a homogeneous dielectric (for right-incident
modes), and the transmitted and reflected components can be derived from the
continuity conditions (\ref{eq:cont}).

For the vector potential of the left-incident mode one finds
\begin{eqnarray}
\label{left}
A^\mu_L(x)&\!\!=&\!\!\! \sum_{\sigma = TE, TM, G, C}\int 
           \frac{d^3k\;\theta(k_z)}{(2\pi)^{3/2}(2\omega)^{1/2} }
 \nonumber\\&&\!\!\!\!\times \bigg\{ e^\mu_\sigma \exp(-ix^0 k^0) a^L_\sigma(k)
 \nonumber \\
  &&\  \times    \Big[\theta(-z)\left(\exp(i{\bf k}\cdot{\bf r}) +
         R^L_\sigma(k) \exp(i{\bf k}^r\cdot{\bf r}) \right) \nonumber\\
  && \ \ \ \ +  \theta(z)\frac{1}{n} T^L_\sigma(k)  
          \exp(i{\bf k}^d\cdot{\bf r}) \Big] + \mbox{h.c.} \bigg\}, 
\end{eqnarray}
where $a^L_\sigma (k)$ is the photon annihilation operator of the mode and
the reflection and transmission coefficients are
\begin{eqnarray}
\label{leftph}
 R^L_{TE}(k)&\!\!=&\!\!\frac{k_z-k_z^d}{k_z+k_z^d}\;,\,\,\,
 T^L_{TE}(k)=\frac{2 n k_z}{k_z+k_z^d}\;,\nonumber \\ 
 R^L_{TM}(k)&\!\!=&\!\!\frac{n^2 k_z-k_z^d}{n^2 k_z+k_z^d}\;,\,\,\,
 T^L_{TM}(k)=\frac{2 n^2 k_z}{n^2 k_z+k_z^d}\;, \nonumber \\  
 R^L_{C}(k) &\!\!=&\!\! R^L_{G}(k)=\frac{k_z- n^2 k_z^d}{k_z+n^2 k_z^d}\;,\nonumber \\
 T^L_{C}(k)&\!\!=&\!\!\frac{2 n^2 k_z}{k_z+ n^2 k_z^d}\;,\,\,\,
 T^L_{G}(k)=\frac{2 n k_z}{k_z+ n^2 k_z^d}\;.
\end{eqnarray}
The wave vectors are
\begin{eqnarray}
\label{k}
 {\bf k} &\!\!=&\!\! ({\bf k}_\parallel, k_z)\;,\,\, 
 {\bf k}^{r}=({\bf k}_\parallel, -k_z)\;,\,\,
 {\bf k}^d = ({\bf k}_\parallel, k_z^d)\;, \,\,\,\nonumber \\ 
 k_z^d &\!\!=&\!\! \mbox{sgn}(k_z)(n^2 k_z^2 + (n^2-1)k_\parallel^2)^{1/2}\;,\,\, 
 k_\parallel^2 = k_x^2 + k_y^2\;,\nonumber \\
 k_0 &\!\!\equiv &\!\! \omega,\,\,\, \omega^2 = k_z^2 +k_\parallel^2 = 
      \frac{1}{n^2}(k_z^{d\,2} +k_\parallel^{2})\;.  
\end{eqnarray}
For the right-incident modes the vector potential reads
\begin{eqnarray}
\label{right}
A^\mu_R(x)&\!\!=&\!\!\! \sum_{\sigma = TE, TM, G, C}\int 
           \frac{d^3k\;\theta(-k_z^d)}{(2\pi)^{3/2}(2\omega)^{1/2} }
 \nonumber\\&&\!\!\!\!\times \bigg\{ e^\mu_\sigma \exp(-ix^0 k^0) a^R_\sigma(k)
 \nonumber \\
  &&\  \times    \Big[\theta(z)\frac{1}{n}\left(\exp(i{\bf k}^d\cdot{\bf r}) +
         R^R_\sigma(k) \exp(i{\bf k}^{r,d}\cdot{\bf r}) \right) \nonumber\\
  && \ \ \ \ +  \theta(-z) T^R_\sigma(k)  
          \exp(i{\bf k}\cdot{\bf r}) \Big] + \mbox{h.c.} \bigg\}, 
\end{eqnarray}
with the reflection and transmission coefficients
\begin{eqnarray}
\label{rightph}
 R^R_{TE}(k)&\!\!=&\!\!\frac{k_z^d-k_z}{k_z+k_z^d}\;,\,\,\,
 T^R_{TE}(k)=\frac{2  k_z^d}{n(k^d_z+k_z)}\;,\nonumber\\
 R^R_{TM}(k)&\!\!=&\!\!\frac{k_z^d -n^2 k_z}{ k_z^d +n^2 k_z}\;,\,\,\,
 T^R_{TM}(k)=\frac{2 k^d_z}{k_z^d + n^2 k_z}\;,\nonumber \\  
 R^R_{C}(k)&\!\!=&\!\! R^R_{G}(k)=\frac{n^2 k_z^d-k_z}{n^2k_z^d+k_z}\;,\nonumber\\
 T^R_{C}(k)&\!\!=&\!\!\frac{2  k_z^d}{n^2k^d_z+k_z}\;,\,\,\,
 T^R_{G}(k)=\frac{2nk_z^d}{n^2k^d_z+k_z}\;.\,\,\,
\end{eqnarray}
Because these modes are right-incident the $z$ component of the incident
wave vector is negative $k^d_z <0 $, and the reflected wave has the wave
vector ${\bf k}^{d,r}=({\bf k}^\parallel,-k^d_z)$. Note that the
integration over ${\bf k}^d $ includes imaginary values of $k_z$, which
correspond to modes that come from inside dielectric, suffer total internal
reflection at the interface, and are evanescent on the vacuum side. One has
\begin{eqnarray}
 k_z &\!\!=&\!\!\mbox{sgn}(k_z^d)\frac{1}{n}\sqrt{k_z^{d\,2} - (n^2 -1)k_\parallel^2}
        \nonumber\\&&\hspace*{30mm}\mbox{for} 
        \ {k_z^{d\,2} - (n^2 -1)k_\parallel^2} >0\;, \nonumber \\        
     &\!\!=&\!\!- i \frac{1}{n}\sqrt{-k_z^{d\,2} + (n^2 -1)k_\parallel^2}
        \nonumber\\&&\hspace*{30mm} \mbox{for}\ {k_z^{d\,2} - (n^2 -1)
	k_\parallel^2} <0\;, \nonumber          
\end{eqnarray}
We have chosen the branch of the square root such that the evanescent modes
are truly evanescent, i.e. exponentially falling away from the interface on
the vacuum side. This also ensures that these modes are genuinely totally 
reflected, i.e. that the relation $ R^R_\sigma {R^R_\sigma}^* =1 $ is
fulfilled.  

For the physical polarizations TE and TM the above modes are well-known
\cite{carn}. It is easy to see that they are mutually orthogonal, but the 
proof of their completeness is surprisingly tricky \cite{BB}. The following
relations are useful for showing the completeness of the modes and for
simplifying our expressions in section \ref{Sec:Wightman}. They are valid
for all polarizations so that we drop the index $\sigma$. For real $k_z$ we
have 
\begin{eqnarray} 
&&\frac{k_z}{k_z^d}n^2 [{T^R}^* T^R](-k_z, -k_z^d) 
+ [{R^L}^* R^L](k_z, k_z^d) = 1 \nonumber\\
&&\frac{k_z^d}{n^2 k_z} [T^L {T^L}^*](k_z, k_z^d) 
+ [R^R {R^R}^*](-k_z,-k_z^d)=1 \nonumber\\
&&\frac{k_z^d}{n^2 k_z} [R^L {T^L}^*](-k_z,-k_z^d) + [T^R {R^R}^*](k_z, k_z^d)
=0\nonumber\\
&&\frac{k_z^d}{n^2 k_z} [{R^L}^* T^L](k_z, k_z^d) + [R^R {T^R}^*](-k_z,-k_z^d)
=0\nonumber\\
&&{R^R}^{*}(-k_z^d) = R^R(k_z^d) \nonumber\\
&&\frac{k_z^d}{n^2 k_z} T^L = T^R\;,\label{eq:relreal}
\end{eqnarray}
and for imaginary $k_z$
\begin{eqnarray}
&&R^R {R^R}^{*} =1\nonumber\\ 
&&{R^R}^{*}(-k_z^d) = R^R(k_z^d) \nonumber\\
&&[{R^R}^* T^R](-k_z^d) = T^R(k_z^d)\nonumber\\ 
&&[R^R {T^R}^*](-k_z^d) = {T^R}^*(k_z^d)\;.\label{eq:relimag} 
\end{eqnarray}

Applying the polarization vectors (\ref{eq:polvecs}) on a plane wave
$e^{i{\bf k}\cdot{\bf r}}$, as in (\ref{left}) and (\ref{right}), one
can express them in terms of the wave vector $k$. However, it is important
to realize that the incident, transmitted, and reflected components all have
different wave vectors and thus, according to (\ref{eq:polvecs}), have
polarization vectors that point in different directions. All four
polarization vectors form a complete and orthogonal system, and the TE and
TM polarizations are complete in the subspace of physical states 
\cite{reinhardt}
\begin{eqnarray}
\label{metrel}
&&   g_{\mu \nu} e^\mu_\sigma e^\nu_{\sigma'} = g_{\sigma \sigma'}, \,\,\,\,
   \sum_{\sigma= TE,TM,C,G} g^{\sigma \sigma} e
   ^\mu_\sigma e^\nu_{\sigma} = g^{\mu \nu}, 
    \nonumber \\
&&   \sum_{\sigma=TE,TM}g^{\sigma \sigma} e^\mu_\sigma e^\nu_{\sigma} = 
   g^{\mu \nu}- \eta^\mu \eta^\nu + {\hat k}^\mu {\hat k}^\nu\;.  
\end{eqnarray}
Here $ {\hat k}^\mu =(k^\mu -(k\eta)\eta^\mu)/\sqrt{(k\eta)^2- k^2}$ is the
unit vector along the space-like part of $k^\mu$, and $\eta^\mu=(1,0,0,0)$.

Finally, we would like to consider the limit $ n \rightarrow \infty $ which
is commonly thought of as corresponding to 
a half space bounded by a perfectly reflecting wall. Indeed, in this limit
the reflection and transmission coefficients become
\begin{eqnarray}
\label{inft}
&&\hspace*{-3mm} R^L_{C}(k)=  R^L_{G}(k) \rightarrow -1,\,\,\,
 R^L_{TE}(k)\rightarrow -1,\,\,\,  R^L_{TM}(k)\rightarrow 1,\,\,\,
   \nonumber \\
&&\hspace*{-3mm} \frac{1}{n} R^R_{\sigma}(k)\rightarrow 0\;,\ 
\frac{1}{n} T^L_{\sigma}(k) \rightarrow 0\;,\ 
 T^R_{\sigma}(k)\rightarrow 0\;,
\end{eqnarray}
so that only the left-incident mode (\ref{left}) survives and gets perfectly
reflected at $z=0$.


\end{document}